\documentclass[prb]{revtex4-1}
\usepackage{dsfont}
\usepackage{amsmath}
\usepackage{enumerate}
\usepackage{amssymb}
\usepackage{graphicx}
\usepackage{wrapfig}
\usepackage{setspace}
\usepackage{cancel}
\usepackage{xcolor}
\def\Xint#1{\mathchoice
   {\XXint\displaystyle\textstyle{#1}}%
   {\XXint\textstyle\scriptstyle{#1}}%
   {\XXint\scriptstyle\scriptscriptstyle{#1}}%
   {\XXint\scriptscriptstyle\scriptscriptstyle{#1}}%
   \!\int}
\def\XXint#1#2#3{{\setbox0=\hbox{$#1{#2#3}{\int}$}
     \vcenter{\hbox{$#2#3$}}\kern-.5\wd0}}

\def\dashint{\Xint-}

\renewcommand{\>}{\rangle}
\newcommand{\<}{\langle}
\newcommand{\tr}{\operatorname{tr}}

\begin{document}

\title{Coherence Factors Beyond the BCS Expressions -- A Derivation}
\author{G. Gorohovsky}
\affiliation{Racah Institute of Physics, Hebrew University,
Jeruslaem, Israel}
\author{E. Bettelheim}
\affiliation{Racah Institute of Physics, Hebrew University,
Jeruslaem, Israel}

\date{\today}

\begin{abstract}
We present a derivation of a previously announced result for matrix elements  between exact eigenstates of the pairing Hamiltnonian. Our results, which generalize the well known BCS (Bardeen-Cooper-Schrieffer) expressions for what is known as 'coherence factors', are derived based on the Slavnov formula for overlaps between Bethe-ansatz states, thus making use of the known connection between the exact diagonalization of the BCS Hamiltonian, due to Richardson, and the algebraic Bethe ansatz. The resulting formula has a compact form after a suitable parameterization of the Energy plane. Although we apply our method here to the pairing Hamiltonian, it may be adjusted to study what is termed the 'Sutherland limit' for exactly solvable models, namely where a macroscopic number of rapidities form a large string.
\end{abstract}

\maketitle

\section{Introduction}

The computation of correlation functions within the Bethe-ansatz approach is an evolving field of study which has seen some notable recent success\cite{Kostov:Three:Point:N4SYM:Announce:PRL,Caux:Calabrese:Slavnov:1P:Dynamical:in:LiebLinger,Calabrese,Kitanine:XXZ:Form:Factors,Biegel:Karbach:Heisenberg}. The computation of correlation functions in this context is often due  to determinant formulas\cite{Slavnov}, which  give overlaps of Bethe-ansatz states in terms of large determinants of matrices, the elements of which are written through  Bethe-ansatz rapidites. Such determinants may then be computed numerically, or  analytically. Here we present a case in the context of condensed matter physics, where a fully analytical  approach is possible, within the thermodynamic limit. In particular, we consider a superconductor described by the pairing Hamiltonian. It should be mentioned that certain eigenstates of the Hamiltonian have a good mean-field description, due to  BCS, which may be used to compute coherence factors. Close to equilibrium these mean field results are enough to describe the physics of a superconductor. Nevertheless, far from equilibrium, eigenstates may be excited which do not have a good mean field description, and for which the most convenient description is in terms of exact solutions. We apply exact methods to compute  between such states. In terms of the Bethe ansatz rapidities, the states which we will consider have two macroscopic strings of rapidities. States described by mean field have a single macroscopic string of rapidities.

The results we obtain were announced  elsewhere \cite{Gorohovsky:Bettelheim:Preparation},  where more discussion on the properties of the solution was provided, while here we provide a derivation.

 \subsection{Richardson's solution for the pairing Hamiltonian}

It is well established that a superconductor may be described by the following effective pairing Hamiltonian:

\begin{align} \label{Hamiltonian}
H = \sum_{j,\sigma_j} \varepsilon_j c^\dagger_{j,\sigma_j} c_{j,\sigma_j} - G \sum_{j,l} c^\dagger_{j,+} c^\dagger_{j,-} c_{l,+}c_{l,-}.
\end{align}
Here $(j,+)$ and $(j,-)$ denote the quantum numbers of time reversed pairs. The indices $j$ and $l$ run from $1$ to $L$ and $\sigma_j$ takes the values $+$ and $-$. For example, if $(j,+)$ denotes a state with wave number $\vec{k}$ and spin up, then $(j,-)$ denotes a state with wave number $-\vec{k}$ and spin down. We assume, for simplicity, that each level $j$ is only doubly degenerate, where $\sigma$ indexes the two degenerate states, $\sigma$ taking $+$ and $-$ as values. Furthermore, we assume uniform level spacing $\varepsilon_j  - \varepsilon_{j-1} = \iota$.

We review now Richardson's solution and then give our expressions for  between exact eigenstates of the Hamiltonian. To describe an exact eigenstate, first note that only pairs of electrons are dynamical, while single electrons decouple from the dynamics, as a result an eigenstate may be described by first specifying the single-particle occupation of the einenstate, namely a set of single particle levels $\{\varepsilon_{j_i}\}_{i=1}^M$, each occupied with an electron with spin $\sigma_j$. A state with such a single-particle occupation, and no pairs is given in the following:

\begin{align}
|\epsilon \>  = \prod_i c^\dagger_{i,\sigma_i} |0\>
\end{align}
An eigenstate is then prescribed by distributing $P$ additional pairs by applying $P$ operators, $b_\alpha^\dagger$, where $\alpha=1,\dots,P$ on the singly occupied state $|\varepsilon\>$
\begin{align}
\label{RichardsonState}
| V,  \epsilon \> = \prod_{\alpha=1}^{P} b^\dagger_\alpha | \epsilon\>.
\end{align}
It remains to give the definition of the operators $b^\dagger_\alpha$. Each of these operators are associated with a rapidity, $v_\alpha$:

\begin{align} \label{PairCreationOperator}
\quad b^\dagger_\alpha = \sum_{ \varepsilon_i \notin \{ \varepsilon_{i_j}\}  } \frac{1}{v_\alpha - \varepsilon_i} c^\dagger_{i, \uparrow} c^\dagger_{i, \downarrow}
\end{align}
The rapidities are subject to Bethe ansatz  equations known as  Richardson's equations. Richardson have found these equations  without recourse to the Bethe ansatz. The Richardson equations state that for each $1 \leq \mu \leq P$ the following equation must be satisfied :
\begin{align}\label{RichardsonEq}
 \sum_{\nu \neq \mu} \frac{1}{ v_\mu-v_\nu}  - \frac{1}{2}\sum_{ \{ i | \nexists \sigma, \varepsilon^\sigma_i \in \epsilon \} } \frac{1}{v_\mu - \varepsilon_i }  - \frac{1}{2g} =0 ,
\end{align}

The energy eigenvalue of a state with rapidities $v_\nu$ and single particle occupancy $\{\varepsilon_{j_i}\}$ is given by $E = \sum_{\nu=1}^P 2 v_\nu + \sum_{ i=1 }^M \varepsilon_{j_i}$.
We shall term a state of the form (\ref{RichardsonState}) a 'Richardson state' even if the rapidities do not satisfy (\ref{RichardsonEq}).  A  Richardson state is, thus, also an eigenstate if and only if the rapidities satisfy (\ref{RichardsonEq}).

The Richardson equations, Eq. (\ref{RichardsonEq}), have a convenient electrostatic interpretation, which will be at the heart of the following development. Define the two dimensional electrostatic field, $h(\xi)$,  associated with the rapidities $V$:
\begin{align} \label{hDefinition}
h^{(V)}(\xi) = \iota\left[ \sum_{\mu=1}^P \frac{1}{\xi-v_\mu}-  \frac{1}{2}\sum_{ \varepsilon_k\notin\{\varepsilon_{j_i}\} } \frac{1}{\xi - \varepsilon_k} - \frac{1}{2g} \right].
\end{align}
Here the rapidities were assigned a charge $1,$ the {\it unblocked} single particle levels (namely those levels which do not contain single electrons)  were assigned a charge $-1/2$ and a background electric field of magnitude $\frac{1}{2g}$ is effected. The Richardson equations demand that all rapidities are  at electrostatic equilibrium.

\begin{figure}[h!!!]
\includegraphics[width=10cm]{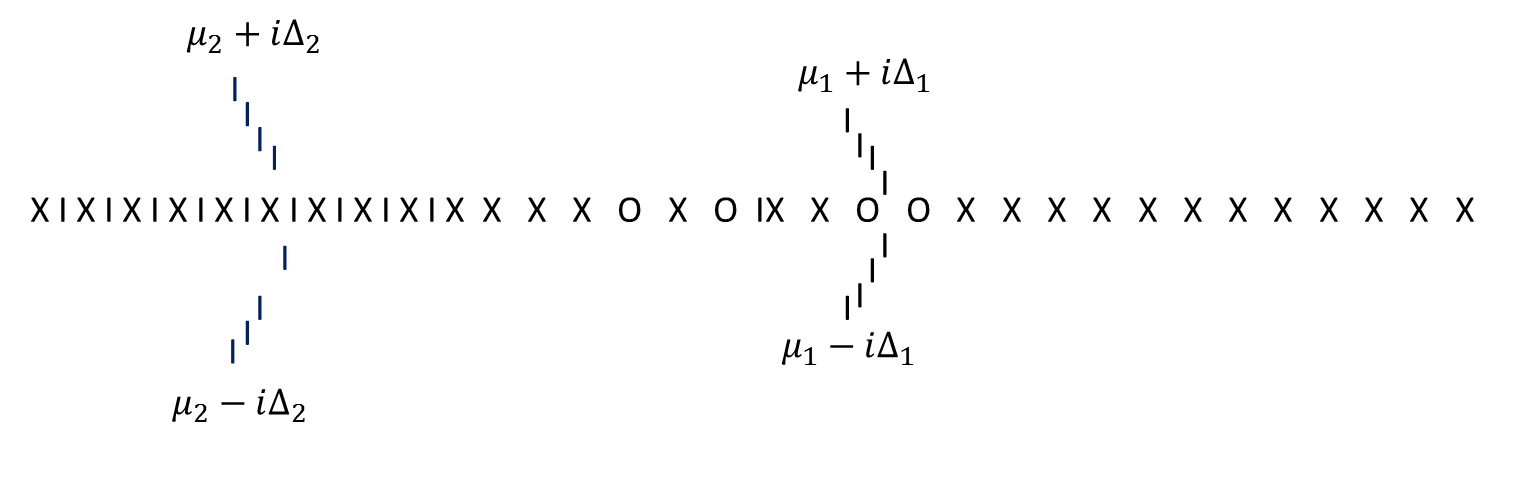}
\caption{A typical configuration of rapidities, the $I$ denote rapidities while the vertical bars, $X$, denote unoccupied single particle levels the $O$'s denote singly occupied  levels.  \label{RichardsonConfigurationFigure}}
\end{figure}

The equations constraining the $v$'s lend themselves to a typical form of the solution. Namely, some of the $v$'s are found on the real axis in more-or-less arbitrary positions dispersed between  the single particle levels, while other $v$'s arrange themselves in arcs in the complex plane. There may be any number of arcs,  while in the present paper we restrict ourselves to the case where there is either $1$ or $2$ arcs. The number of arcs will be denoted by $k$, and the end points of the arcs by  $\{\mu_{i}\pm \Delta_{i}\}_{i=1}^k$. Fig. \ref{RichardsonConfigurationFigure} displays such a typical configuration with two arcs.

In the thermodynamic limit we describe the distribution of single particle levels, singly occupied states and the rapidities by coarse-grained densities. We multiply all densities by the level spacing to obtain   'occupancy numbers'.  While the density of the $v$'s on the real axis is largely arbitrary, the density on the arcs may be found given the real-axis density of $v_\mu$'s and $\varepsilon_{i_j}$'s and the arc endpoints $\{\mu_{i}\pm \Delta_{i}\}_{i=1}^k$. We may then compute the  between two Richardson states each described by its real-axis density and its arc end-point. We denote such a state by $|n_+,n_-,n_V,\{\mu_{i}, \Delta_{i}\}_{i=1}^k\>.$ Here and below  $n_\alpha$, where $\alpha $ an take the 'values' $+, -$ or $V$ are  given in the following:
$n_\alpha(\varepsilon)= \frac{\iota}{\delta \varepsilon}\Xi_\alpha(\varepsilon) ,
$
where $\Xi_V(\varepsilon)$, $\Xi_{+}(\varepsilon)$ and $\Xi_{-}(\varepsilon)$ are respectively the number of rapidities, singly occupied levels with spin $+$ and singly occupied levels with spin $-$ in the segment $[\varepsilon-\frac{\delta \epsilon}{2},\varepsilon+\frac{\delta \epsilon}{2}]$ and $\delta \varepsilon$ is a coarse graining scale defined such that it is much larger than $\iota$, the level spacing, and much smaller than the scale at which densities change. We  define an excitation occupation number, $n(\varepsilon),$ as follows:
\begin{align}
n(\varepsilon)=n_+(\varepsilon) + n_-(\varepsilon) + 2n_V(\varepsilon).\label{nDef}
\end{align}

 The field $h^{(V)}(\xi)$ consequently has a jump discontinuity on the real axis of magnitude $\pi i\left({n(\varepsilon)-1}\right) $ and on the arcs, where it has some $O(1)$ jump discontinuity, which must be determined.  Consider the endpoints of the arcs. Those rapidities that lie on the endpoints, must be in electrostatic equilibrium. A closer analysis shows that this is only possible if the field $h^{(V)}(\xi)$ as $\xi$ approaches the endpoint tends to $0$. This is an intuitive result, since if $h^{(V)}(\xi)$ would not approach $0$ the endpoints would feel a force that would move them. One concludes that $h^{(V)}(\xi)$ vanishes on the 2K endpoints of the arcs. Moreover, if we look at the average value of $h^{(V)}(\xi)$ across the arc (namely $\frac{h^{(V)}(\xi_+)+h^{(V)}(\xi_-)}{2}$, where $\xi_\pm$ are points just to the left and to the right of the arc respectively), then this average must vanish. The reason being, that this average represents the far-field felt by the charges on the arc. Looking under a magnifying glass at a segment of the arc, one sees a long (from this perspective, infinite) chain of charges. These chains will fly off if an external (or far-) field is present. Namely, the average must vanish. These considerations allow finding $h^{(V)}(\xi)$. In fact Gaudin in a paper in French\cite{Gaudin} (later reviewed and expanded in English in Ref. [\onlinecite{Sierra}]) had shown that it is given by
\begin{align} \label{hIntegral}
h^{(V)}(\xi) = R_{2K}(\xi) \int \frac{ n(\varepsilon)-1}{2R_{2K}(\varepsilon)(\varepsilon-\xi)} d\varepsilon,
\end{align}
where
\begin{align}\label{Rdefinition}
R_{2K}(\xi) = \prod_{j=1}^K \sqrt{(\xi - \mu_j)^2 +\Delta_j^2}
\end{align}
Indeed, $h^{(V)}(\xi)$ defined by (\ref{hIntegral})  has a jump continuity on the real axis of the given value $\pi i \left({n(\varepsilon)-1}\right) $, vanishes on the endpoints of the arc has a non-trivial jump discontinuity on  $K$ arcs, and  its average value across an arc is $0$ (since it simply changes sign across the arc).

Eq. (\ref{hIntegral}) is an expression for $h^{(V)}(\xi)$ given a knowledge of $n(\varepsilon)$ and of the endpoints of the arcs $\mu_j, \Delta_j$, $j=1,\dots,K$. $n(\varepsilon)$ is arbitrary and may be tuned by  blocking levels and placing rapidities between adjacent unblocked levels. The arc endpoints must be determined self-consistently, however. These self-consistency conditions can be derived by noting that $h^{(V)}(\xi)$ as defined by (\ref{hDefinition})  must have the following asymptotic behavior as $\xi\to\infty$: $h^{(V)}(\xi) \to \frac{1}{2g} +O\left(\frac{1}{\xi} \right)$. Expanding in large  $\xi$, Eq. (\ref{hIntegral}) shows that the expected asymptotic behavior of $h^{(V)}(\xi)$ is only satisfied if the following $K-1$ conditions hold:
\begin{align}\label{selfconsistencies}
\int \frac{n(\varepsilon) \varepsilon^l }{R_{2K}(\varepsilon)} d\varepsilon = \frac{1}{g}\delta_{l,K-1} , \quad l\leq K-1.
\end{align}
These are not enough to determine $2K$ free parameters which determine the position of the endpoints. Extra conditions may be found if one knows the number of rapidities on each one of the arcs. In the case of one arc, the number of rapidities on the arc is known if one knows the total number of electrons. Indeed, the total number of particles is $2P+M$, where $P$ is the number of rapidities and $M$ are the number of singly occupying electrons. The number of rapidities on the real axis is known since we know $n(\varepsilon)$ and so the number of rapidities on the arc is also known. If there is more than one arc, however, the total number of particles is not enough to fully determine the endpoints, and for the same number of particles, the same given $n(\varepsilon)$ and the same number of arcs, one may find different solutions depending on how many rapidities occupy each arc. The solutions differ by the location of the endpoints of the arcs.

The continuum limit is achieved by taking the level spacing $\iota$ while for any given finite energy window $[\xi - \Delta \xi , \xi+ \Delta \xi]$ the ratio between the number of single particle levels in the window, which is given by $\frac{\Delta \xi}{\iota}$, and the number of rapidities is finite. The same is true for the ratio between the number of singly occupied levels and the number of single particle levels. The number of pairs also scales with $\iota$, namely $P\iota \to \rm{const}$  in the thermodynamic limit.

\subsection{Abel Map}
\begin{figure}[h!!!]
\includegraphics[width=3cm]{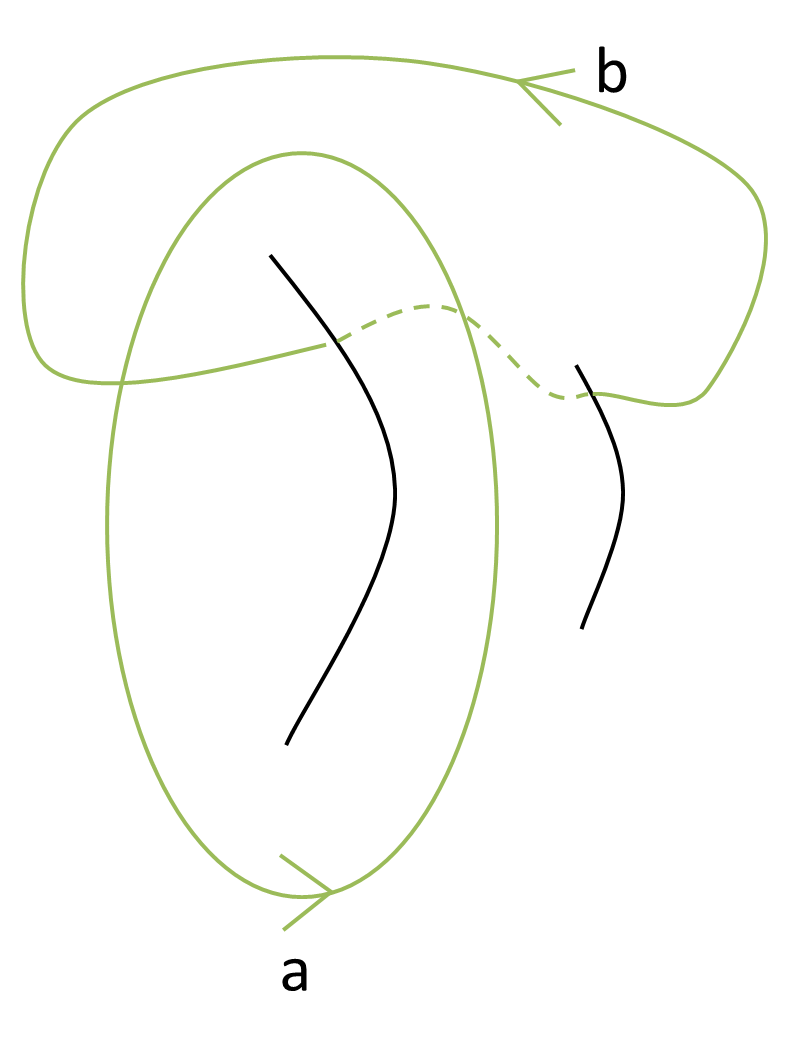}
\caption{$a$ and $b$ cycles on the algebraic Riemann surface defined by $\sqrt{R_4(\xi)}$. The two branch cuts are drawn in black. The $a$ and $b$ cycles are drawn in gray. The portion of the  $b$ cycle  on the lower sheet is drawn as a dashed line. \label{Cycles}}
\end{figure}

The appearance of the function $R_4(\xi)$ in the equations determining the electrostatic field, the self-consistency equations, etc., establishes a connection between the problem at hand and the theory of algebraic Riemann surfaces. Indeed, the function $R_4(\xi)$ , which is multi-valued, becomes single-valued if one takes two copies of the complex plane, assign one branch of the function $R_4(\xi)$ to each one of the copies and glues these two sheets along the branch cuts of the function $R_4(\xi)$. We denote the resulting Riemann surface by $\mathcal{R}$. The topology   of $\mathcal{R}$ is that of a torus. A torus has two non-trivial cycles, two closed contours the cannot be deformed into one another or shrunk to a point. These two cycles are standardly  denoted by $a$ and $b$ (see Fig. \ref{Cycles}).

The Abel map takes the two sheeted Riemann surface and maps it one to one  and onto a rectangle  in which opposite sides are identified (the latter being  a popular  representation of the torus). The map reads explicitly:
\begin{align}\label{AbelMap}
u_\xi =\frac{\pi\int_{\mu_1+i \Delta_1}^\xi \frac{1}{R_{4}(\xi')} d\xi}{ 2\int_{\mu_1+i \Delta_1}^{\mu_1-i \Delta_1} \frac{1}{R_{4}(\xi')} d\xi}.
\end{align}
Note that there are many paths of integration one may choose  for each given $\xi$. For example for any given path one may obtain another path by adding any number $m$ and $n$ of cycles $a$ and $b$, respectively. This is achieved by letting the path wind along the cycles $a$ and $b$ before reaching $\xi$. Adding $m$ cycles of type $a$ and $n$ cycles of type $b$ results in shifting the image $u_\xi$ by $m\pi+n\tau$, where the purely imaginary number $\tau$ is defined by
\begin{align}
\tau =2u_{\mu_2+i \Delta_2}.\label{tauDef}
\end{align}
To obtain a well defined map, we must identify points shifted by $m\pi+n\tau$. So that the Abel map is a function from the Riemann surface, $\mathcal{R}$, to $\mathbb{C}/\mathbb{Z}\pi+\mathbb{Z}\tau$. The unit cell of the lattice $\mathbb{Z}\pi+\mathbb{Z}\tau$ is a rectangle with sides of length  $\pi$ and $|\tau|$ and the equivalence identifies points on opposite sides, so that we may equivalently view the map $u_\xi$ as mapping $\mathcal{R}$ onto the this rectangle.

\subsection{Results}
The Abel maps provide a convenient representation in which we may write our results.  

The state  $c^\dagger_{m\sigma}|n_+,n_-,n_V,\{\mu_{i}, \Delta_{i}\}_{i=1}^k\>$  will generally  have little overlap with states with significantly different occupation numbers, $n_\alpha$ and arc-endpoints. Thus denoting the in-state, $|\mbox{in}\> $, by $|n_+,n_-,n_V,\{\mu_{i}, \Delta_{i}\}_{i=1}^k\>,$ we compute the matrix element $\<\mbox{out}|c^\dagger_{m,\sigma}|\mbox{in}\>$, where the state  $|\mbox{out}\>$ may be considered to have the same density and arc-endpoints as  $|\mbox{in}\>$. Nevertheless, the object, $\<\mbox{out}|c^\dagger_{m,\sigma}|\mbox{in}\>,$ is not a diagonal matrix element since $|\mbox{out}\>$ may be different from the in-state on a microscopic scale.

We describe the difference between $|\rm{in}\>$ and $\<\rm{out}|$ states  by two variables, $p$ and $l$. We claim that all order $1$ overlaps  are covered by the following values of $p$ and $l$.  The number  $p$ is any integer much smaller than $N$ counting how many more rapidities are on the left arc in the $|\mbox{out}\>$  state as compared to the $|\mbox{in}\>$ state. The number  $l$ is defined to be $1$ $(-1)$ if the $|\mbox{out}\>$ state has one excitation more (less)  next to $\varepsilon_m$ as compared to $|\mbox{in}\>$. Note that $c^\dagger$ ostensibly creates an excitation, so  naively $l=1$, however, a well known feature of superconductivity is that a condensation of a pair may accompany the creation of an excitation.   The latter corresponds here to a rapidity leaving the vicinity of $\varepsilon_m$ and joining an arc -- a process which brings $l$ down to $-1$.

\begin{figure}[h!!!]
\includegraphics[height=5cm]{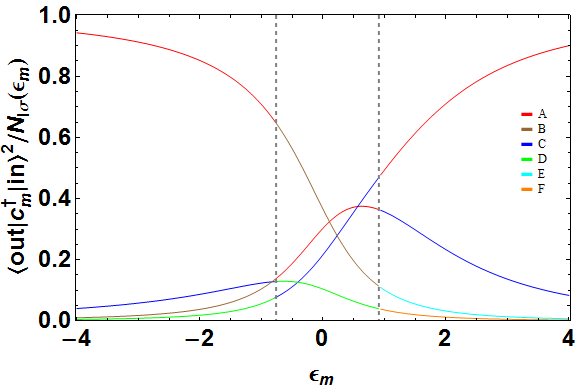}
\caption{
\label{ResultGraph}
 Matrix element square between an in- and out-state divided by $N_{l\sigma}$. Each color corresponds to a different out-state according to the following key.
{\bf A}: l=1,p=0;
{\bf B}: l=-1,p=1;
{\bf C}: l=-1,p=0;
{\bf D}: l=1,p=-1;
{\bf E}: l=1,p=1;
{\bf F}: l=-1,p=-1;
 The states correspond to $\Delta_1=2.2534$, $\mu_1=0.8257$, $\Delta_2=1.0612$ and $\mu_2=-0.0499$. }

\end{figure}

We  now define:
\begin{align}
N_{l,\sigma}=\delta_{l,1}-l\left(n_{l\sigma}+n_V\right)
\end{align}
allowing us to formulate  the main result of the paper:

\begin{align}\label{IntroductionResults}
&\<\mbox{in};l,p|c^\dagger_{m\sigma}|\mbox{in}\>^2=\frac{\pi^2 l N_{l,\sigma}\sin^{-2}\left[u_{\varepsilon_m}+l(p\tau-u_\infty)\right]  }{4\omega^2 R_4({\varepsilon_m})}
\end{align}
where
\begin{align}
& \qquad \omega =\int_{\mu_1-i \Delta_1}^{\mu_2+i \Delta_1} \frac{1}{R_{4}(\xi')} d\xi
\end{align}
In Fig. \ref{ResultGraph} the matrix element square is drawn for a few values of $l$ and $p$. 

\section{Coherence Factors Through Slavnov Overlaps\label{ThroughSlavnovSetion}}
We wish to compute the matrix element of a single fermionic creation operator between two eigenstates of the Hamiltonian.
Our approach is to make of use Slavnov's formula, which gives the overlap between two Richardson states
 in terms of a determinant, which in turn is written in terms of the rapidities of the two states. Slavnov's formula assumes that one of the states in question is an eigenstate, while the other is not (if both are eigenstates then the result is either $0$, if the states are different or, otherwise,  the norm of the state).
 Our first goal is then to write the  as Such overlaps.

We note that that matrix elements the  of any number of fermionic operators may be computed through the matrix elements of single fermionic operators by the insertion of a complete set of states. We thus concentrate only on the computation of coherence factors of single fermionic operators. There are however several cases we have to consider. First we may consider matrix elements of  of $c_{m,\sigma}$ and $c^\dagger_{m,\sigma}$. Second we may consider matrix elements where the single particle level, $\varepsilon_m$, on which $c^\dagger_{m\sigma}$ or  $c_{m\sigma}$ act may be either singly occupied or not. One may develop a Slavnov overlap approach for each one of these four cases, but this is not necessary since all four possibilites may be deduced from one, making use of simple symmetries\cite{Pogosov:Electron:Hole:Richardson}. We digress to describe this reduction to one type of matrix element  (in particular the matrix element of $c^\dagger_{m\sigma}$ where $\varepsilon_m$ is not ingly occupied) in the next subsection before writing it in terms of Slavnov overlaps in the following section.

\subsection{Reduction to One type of Matrix Element\label{OneMatrixElemSection}}

First note that taking the complex conjugate of  the result, Eq. (\ref{IntroductionResults}), will give the  of $c_{m,\sigma}$ with the interchange of the in and out states, which amounts to taking $l\to-l$ and $p\to-p,$ thus we have:
\begin{align}
\<\mbox{in};l,p|c_{m\sigma}|\mbox{in}\>^2=\frac{-\pi^2 l N_{-l,\sigma}\sin^{-2}\left[u_{\varepsilon_m}+l(p\tau+ u_\infty)\right]  }{4\omega^2 R_4({\varepsilon_m})}.
\end{align}
Accordingly, we need not worry about computing the  between of the fermionic annihilation operator and concentrate solely on the creation operator.

We now show that we may also infer the matrix element when the single particle level $\varepsilon_{m}$ is not singly occupied from the case where the single particle level is  singly occupied, allowing us to make all computations assuming the latter. To see this consider that the level $\varepsilon_m$ is  singly occupied and consider the matrix element
$
\<{\rm B}|c^\dagger_{m,\sigma}|{\rm A}\>
$. Note that the level $\varepsilon_m$ is not singly occupied in the $\<{\rm B}|$ state. We make use of the fact that the Richardson state is written in a gauge where the wave function is real to obtain $
\<{\rm B}|c^\dagger_{m,\sigma}|{\rm A}\> = \<{\rm A}|c_{m,\sigma}|{\rm B}\>
$.  The matrix element  $
 \<{\rm A}|c_{m,\sigma}|{\rm B}\>
$ has the advantage that the in state, $|{\rm B}\>$ , is not singly occupied at level $\varepsilon_m$ at the price of it being a matrix element of a fermionic annihilation operator, rather than a creation operator. Nevertheless, we may use particle-hole symmetery   (discussed with regards the Richardson solution in Ref.  [\onlinecite{Pogosov:Electron:Hole:Richardson}]) to relate  $
 \<{\rm A}|c_{m,\sigma}|{\rm B}\>
$     to  $
 \<{\rm A}|c^{\dagger}_{m,\sigma}|{\rm B}\>.
$ Indeed the Hamiltonian, Eq. (\ref{Hamiltonian}), which we write as  $H(\{c_{m,\sigma}\},\{c^\dagger_{m,\sigma}\};G )$ to  denote it as an operator  function of the creation and annihilation operators and the coupling constant enjoys the symmetry:
\begin{align}
\label{PHSymmetry}H(\{c_{m,\sigma}\},\{c^\dagger_{m,\sigma}\};G )=-H(\{c^\dagger_{m,\sigma}\},\{c_{m,\sigma}\};-G),
\end{align}
which will allow us to relate  of fermionic creation and annihilation operators, respectively.

Both the Hamiltnonian on the  right hand side and left hand side of (\ref{PHSymmetry})
may be solved using Richardson's approach. We are already familiar with the solution of the Hamiltonian on the right hand side of this equation, where the vacuum where no electrons are present is filled up first with single electrons and then with pairs. On the right hand side we may use Richardson's approach by taking a vacuum where no holes are present and fill it up first with single holes and then pairs of holes. In the procedure based on the particle picture (namely that based on diagonalizing the Hamiltonian on the left hand side of (\ref{PHSymmetry})) the eigenstates are labeled by $\Delta_i$, $\mu_i$,  $n_\sigma$ and $n_V$, while the procedure based on holes will yield eigenstates which we label by  $\tilde\Delta_i$, $\tilde \mu_i$,  $\tilde n_\sigma$ and $\tilde n_V.$   Due to the equality between right and left hand sides in (\ref{PHSymmetry}) the set of eigenstates which each procedure yields must be the same. The mapping between the set of eigenstates produced  in the particle picture and the one produced within the hole picture is as follows:
\begin{align}
&\tilde{\Delta}_i = \Delta_i, \quad \tilde \mu_i =\mu_i \\
\label{PHns} & \tilde{n}_\sigma = n_{-\sigma}, \quad \tilde n_V = 1-n_V-\Sigma _\sigma n _\sigma,
\end{align}
where this result may be obtained by considering that the excitation spectrum around each solution, the expectation values, etc. must be the same for both representation of the same eigenfunction, and then comparing the results for the particle and hole representations.

We denote now the number of rapidities on each of the arcs by $N_i$, with $\tilde{N}_i$ being the number of rapidities on the arcs in the hole picture. One may compute $N_i$\ by writing it as a proper contour integral over $h$ around the respective arcs. Making use of this contour integral representation, the following relation between the two may be deduced $\tilde N_i = C - N_i$, where $C_i$ is a constant given by:
\begin{align}
C_i=\int_{\alpha_i}^{\tilde \alpha_i} n(\varepsilon) d \varepsilon
\end{align}
and  $\alpha_i$ and $\tilde \alpha_i$ is the location where arc $i$ cuts the real axis in the particle and hole representation, respectively. In principle, $\alpha_i$ differs from $\tilde \alpha_i$, other wise $C_i$ would be zero and we would get the paradoxical result that the number of rapidities in the hole picture, $\tilde{N}_i$, is negative. Nevertheless, in practice, and quite formally, one may choose $\alpha_i = \tilde \alpha_i$, and work with  $\tilde N_i = -N_i. $ The latter choice is more convenient, since otherwise Eq. (\ref{PHns}) stated above as fact, must be modified in the region between $\alpha_i$ and $\tilde \alpha_i$.

We wish to use this correspondence between the particle and hole representation to relate $
\<{\rm B}|c^\dagger_{m,\sigma}|{\rm A}\>$ to a matrix element where $c^\dagger_{m,\sigma}$ acts on a state with no single electron at $\varepsilon_m$. Suppose the state $|{\rm A}\>$ corresponds to given values of  $n^0_\sigma$, $n^0_V,$ $\Delta^0_i$ and $\mu^0_i$ and the matrix element corresponds to some values  $p=p^0$ and $l=l^0$. We have already mentioned that this matrix element is equal to   $
 \<{\rm A}|c_{m,\sigma}|{\rm B}\>
$, which can be considered as the matrix element of a particle annihilation operator acting on a state with $n^0_\sigma$, $n^0_V,$ $\Delta^0_i$ and $\mu^0_i$    with the values $p=-p^0$ and $l=-l^0$. Alternatively, we may interpret this matrix element as a {\it hole} creation operator with spin $-\sigma $ acting on a state with  $\tilde n_\sigma =-n^0_\sigma$, $\tilde n_V=1-n^0_V-\sum_\sigma n^0_\sigma,$ $\tilde \Delta_i = \Delta^0_i$ and $\tilde \mu_i =\mu^0_i$ and values of $p$\ and $l$ given by $p=p_0,$ $   l=l_0$.

The conclusion we draw is that the matrix element $\<\mbox{in};l,p|c^\dagger_{m\sigma}|\mbox{in}\>$ where the single particle level $\varepsilon_m$ is  singly occupied is the same as $\<\mbox{in};l,p|c^\dagger_{m\sigma}|\mbox{in}\> $ where the state $\varepsilon_m$ is not singly occupied but with a different configuration of rapidities and singly occupied levels on the real axis, but with the same $l$, $p$, $\Delta_i$ and $\mu_i$.

\subsection{Matrix Element in Terms of Slavnov Overlaps}

We now set $|{\rm in}\> = \frac{|W,\epsilon^w\>}{ \sqrt{ \< W, \epsilon^w  |   W, \epsilon^w \>}}  $, namely the $|{\rm in}\>$ state is a normalized Richardson state with rapidities $w_i$ and single occupation content $\epsilon^w$.  Similarly, the out state is given  by  $|{\rm out}\> = \frac{|V,\epsilon^v\>}{ \sqrt{ \< V, \epsilon^v  |   V, \epsilon^v \>}}  $ . Our aim is to  compute
\begin{align}\label{rasSwTilde}
& \<{\rm out}|c^\dagger_{m\sigma}|{\rm in}\>= \frac{ \< V, \epsilon^v | c_{m,\sigma}^\dagger |  W, \epsilon^w  \> }{\sqrt{ \< V, \epsilon^v  |   V, \epsilon^v \> } \sqrt{ \< W, \epsilon^w  |   W, \epsilon^w \> }}.
\end{align}

The numerator can be represented in two different ways as an overlap between two Richardson state, where only one of the states obeys the Richardson equations. As mentioned above (section \ref{OneMatrixElemSection}), without loss of generality, we may assume that $ \varepsilon^{\bar{\sigma}}_m, \varepsilon^{\sigma}_m \notin \epsilon^w$. The state that is created by $c^\dagger_{m,\sigma} |W ,  \epsilon^w \>$ will necessarily have a single electron with spin $\sigma$ occupying the single particle level $\varepsilon_m$, and an analysis of the occupation of all unblocked single particle levels shows that they do not lose the Richardson form. This conclusion can be put in mathematical form:
\begin{align}
c^\dagger_{m,\sigma} | W, \epsilon^w  \> = |W, \epsilon^w \cup \{\varepsilon^\sigma_m\}\>
\end{align}
In fact, $\epsilon^w \cup \{\varepsilon^\sigma_m\} =\epsilon^v$, so that we can write (\ref{rasSwTilde}) as:
\begin{align}\label{OverlapRep1}
&\<{\rm out}|c^\dagger_{m\sigma}|{\rm in}\>= \frac{ \< V, \epsilon^v |W, \epsilon^v  \> }{\sqrt{ \< V, \epsilon^v  |   V, \epsilon^v \> } \sqrt{ \< W, \epsilon^w  |   W, \epsilon^w \> }},
\end{align}
which is a desired representation of the  matrix element in term of Slavnov overlaps.

The representation (\ref{OverlapRep1}) is in fact not very convenient since it distinguishes $w$ over $v$. To obtain the object in a more convenient form we now consider $\< V, \epsilon^v | c_{m,\sigma}^\dagger$.
We may write:
\begin{align}\label{cdaggerFromRightAsTilde}
\<V, \epsilon^v |c_{m,\sigma}^\dagger =\<V,\epsilon^w|-\sum_\alpha \frac{1}{v_\alpha - \varepsilon_m} \<V\setminus\{v_\alpha\} , \epsilon^w  |c_{m+}c_{m-}
\end{align}
Here after the action of $c^\dagger_{m,\sigma}$ the state  $\<V, \epsilon^v |$ loses the single electron at $\varepsilon_m$ such that the single occupation  particle content is the same as $\epsilon^w$, and as such all states on the right hand side have $\epsilon^w$ as a label. In addition there is no pair at $\varepsilon_m$ in $\<V, \epsilon^v |c_{m,\sigma}^\dagger$. This is achieved by removing from $\<V,\epsilon^w|$ all the possible ways in which the level $\varepsilon_m$ may have been occupied by a pair. Namely, one should remove the contribution of each of the pair creation operator $b^\dagger_\alpha$. This is the done by the sum on the right hand side, where the summand is the quantum state given that operator $b^\dagger_\alpha$ creates a pair at $\varepsilon_m$. After this heuristic motivation it is easy to check through the definition of the Richardson state, Eq. (\ref{RichardsonState}), that Eq. (\ref{cdaggerFromRightAsTilde}) holds. We may further rewrite (\ref{cdaggerFromRightAsTilde}) by noting that the pair created at $\varepsilon_m$ by the operator $c_{m+}c_{m-}$ (acting from the right) may be alternatively created by bringing a rapidity close to $\epsilon_m$: 
\begin{align}
\<V, \epsilon^v |c_{m,\sigma}^\dagger =\<V,\epsilon^w|-\lim_{\delta \to 0}\sum_\alpha \delta \frac{\<V\setminus\{v_\alpha\}\cup\{\epsilon_m+\delta\} , \epsilon^w  |}{v_\alpha - \varepsilon_m}
\end{align}
We arrive at:
\begin{align}
\label{OverlapRep2}\<{\rm out}|c^\dagger_{m,\sigma}|{\rm in}\> =\frac{\<V,\epsilon^w|W,\epsilon^w\>}{\sqrt{\<V, \epsilon^v|V, \epsilon^v\>}\sqrt{\<W, \epsilon^w|W, \epsilon^w\>}} \left( 1-\lim_{\delta\to0 }\sum_\alpha \delta \frac{\<V\setminus\{v_\alpha\}\cup\{\epsilon_m+\delta\} , \epsilon^w  |W,\epsilon^w\>}{(v_\alpha - \varepsilon_m)\<V,\epsilon^w|W,\epsilon^w\>}\right)
\end{align}

We can combine the representation (\ref{OverlapRep1}) and (\ref{OverlapRep2}) into one by taking the geometric average of the two:
\begin{align}\label{rThroughOverlaps}
r(m^\sigma ; V, \epsilon^v,W, \epsilon^w) = \frac{\sqrt{\<V, \epsilon^v|{W},{\epsilon}^v\>} \sqrt{\<V {\epsilon}^w| W, \epsilon^w\>}}{ \sqrt{\<W, \epsilon^w|W, \epsilon^w\>} \sqrt{\<V, \epsilon^v|V, \epsilon^v\>}} \sqrt{ 1-\lim_{\delta\to0 }\sum_\alpha \delta \frac{\<V\setminus\{v_\alpha\}\cup\{\epsilon_m+\delta\} , \epsilon^w  |W,\epsilon^w\>}{(v_\alpha - \varepsilon_m)\<V,\epsilon^w|W,\epsilon^w\>}}.
\end{align}
The matrix element is thus re-written entirely through Slavnov overlaps.

\subsection{Slavnov's Formula \label{SlavnovFormula}}
We now present  Slavnov's formula\cite{Slavnov}, which will allow us to compute (\ref{rThroughOverlaps}). We shall also re-write it in a form more suitable to our method.  Slavnov's formula being applied\cite{Zhou:Links:McKenzie:Gould} to the Richardson solution reads:
\begin{align}\label{Slavnov}
\< V, \epsilon | W, \epsilon \>  = \frac{\prod_{a, b} (v_b - w_a )}{\prod_{b<a} (w_b-w_a) \prod_{a<b}(v_b-v_a)} \det J,
\end{align}
 The set $V = \{v_\nu\}_{\nu=1}^P$  obeys the Richardson equations, while $W = \{w_\nu\}_{\nu=1}^P$   do not necessarily satisfy the Richardson equations. Both states share the same single particle occupancy $\epsilon =\{\varepsilon_{j_i}\}$ (otherwise the overlap is $0$).  The matrix $J$ appearing in (\ref{Slavnov}) is given by:

\begin{align}\label{JDefinition}
J_{ab} = \frac{1}{v_a-w_b} \left(\sum_{i=1}^M \frac{1}{(v_a-\varepsilon_{j_i})(w_b-\varepsilon_{j_i})} -2 \sum_{c\neq a} \frac{1}{(v_a-v_c)(w_b-v_c)} \right)
\end{align}

We shall want to write $J_{ab}$ in a more convenient form for our purposes. Let us quantify the deviation of the set $W$ from being a solution of the Richardson equations by the field $\Delta H^{W,\epsilon}(\xi)$, a meromorphic functions satisfying:
\begin{align}
\Delta H^{W,\epsilon}(w_a) = \sum_{b\neq a} \frac{1}{w_a -  w_b} -\frac{1}{2}\sum_{\varepsilon_i\in \epsilon} \frac{1}{w_a - \varepsilon_i} -\frac{1}{2g}\label{DeltaHDef}.
\end{align}
The latter equation does not fix $\Delta H^{W,\epsilon}(\xi)$ uniquely, as it only gives its value at the point $w_a$. We shall see, however that in all relevant cases there is a convenient form for $\Delta H^{W,\epsilon}(w_a)$, which is natural. For example,  if the set $W$ satisfies the Richardson equations, then we may choose $\Delta H^{W,\epsilon}(\xi)=0$, as a convenient form.

With this definition $J_{ab}$ may be recast in the form:
\begin{align}
J_{ab} = \frac{2}{(w_b -v_a)^2}\left(\delta H(w_b) -  \Delta H^{W,\epsilon}(w_b) + \frac{1}{w_b-v_a} \right)
\end{align}
where
\begin{align}\label{BigdeltaHDef}
\delta H(w_b) = \sum_{c \neq b} \frac{1}{w_b -  w_c}  -  \sum_{c } \frac{1}{w_b -  v_c}   ,
\end{align}
We may further simplify the representation of overlaps as a determinant by multiplying it by the Cauchy matrix:
\begin{align}
\label{CauchyC}C_{ab} = \frac{1}{w_a -v_b}
\end{align}
and a diagonal matrix, $\Lambda$.  Where the latter is defined by:
\begin{align}
\label{LabmbdaMat}\Lambda_{ab} =  \delta_{ab}  \frac{\prod_c v_a-w_c}{\prod_{ c \neq a} v_a-v_c}  .
\end{align}
Defining
\begin{align}
F(x) = \frac{\prod_c x-w_c}{\prod_{ c } x-v_c} ,\quad  \chi_a (x) = \frac{\prod_{c\neq a} x- w_c}{\prod_c x-v_c}
\end{align}
The following identity holds:
\begin{align}\label{MatrixProduct}
&\left[ C \Lambda J \right]_{ab} = \oint_V F(x) \frac{2}{(x-w_a)(x-w_b)^2} \left(\delta H(w_b) -  \Delta H^{W,\epsilon}(w_b) + \frac{1}{w_b-x} \right) dx = \nonumber \\
& = \left\{ \begin{array}{lr}
\frac{ 2\chi_b (w_b)}{(w_b-w_a)}\left[ \frac{1}{w_b-w_a}-  \Delta H^{W,\epsilon}(w_b) \right], & a\neq b\\
\quad & \quad \\
\chi_b (w_b)\left[\delta H(w_b)^2 -\delta H'(w_b)  - 2 \delta H(w_b) \Delta H ^{W,\epsilon}(w_b) \right], & a= b
\end{array} \right. ,
\end{align}
where the contour integral is taken to surround the points $v_c$. The first equality is due to picking up the residues at $v_c$, while the second equality involves deforming the contour to surround the point $w_a$ and $w_b$ instead, and picking up the residues there.

Since the determinant of the matrix $C$ is given by the Cauchy formula:
\begin{align}
\det (C) = \frac{\prod_{b<a} (w_b-w_a) \prod_{a<b}(v_b-v_a)}{\prod_{a, b} (v_b - w_a )},
\end{align}
one obtains the following formula for the overlap:
\begin{align}\label{OverlapsAsDdeterminant}
&\< W, \epsilon |V, \epsilon \>  =  2^P\det D(W,V;\epsilon)  \nonumber\\
&D(W,V;\epsilon)_{ab} = \left\{ \begin{array}{lcr}
\frac{ \delta H(w_b)^2 -\delta H'(w_b) }{2} - \delta H(w_b) \Delta H^{W,\epsilon} (w_b), &\quad & a=b \\
 \frac{1}{(w_b-w_a)^2} -  \frac{ \Delta H^{W,\epsilon}(w_a) }{w_a-w_b}, &\quad & a\neq b
\end{array}\right. .
\end{align}

When the set $V$ coincides with the set $W$, the manipulations above cannot be carried through. In this case, using Slavnov's formula directly gives the norm of the Richardson state.
\begin{align}
\<W , \epsilon  | W, \epsilon\> =2^P \det \left(N(W) \right)
\end{align}
\begin{align}\label{NDefinition}
N(W)_{ab} = \left\{ \begin{array}{lr}
\iota^{-1}h'^{(W)}(w_b),  & a=b \\
\frac{1}{(w_b-w_a)^2},  & a\neq b
\end{array} \right. ,
\end{align}
where we indicate explicitly the dependence of the matrix $N(W)$ on the rapidities $W$.  This limit form, $N(W)$, of $J$ will appear frequently in the sequel.
\subsection{Operational Form of the Coherence Factors}
We now make the simplifying assumption that no rapidities in the set  $W$ (or, respectively, $V$) for $l=1$ (respectively, $l=-1$) in a region around $\varepsilon_m$ of large microscopic size, namely,  a region of size $K \iota$, where $K\gg 1$ and $K$ does not scale with $N$ (in the thermodynamic limit $K/N\to0$).    Note that for $l=-1$ the simplifying assumption implies that there is one rapidity, which we assume to be the 'last' rapidity,  $w_P$, in the region around $\varepsilon_m$.

The equations for the  in terms of Slavnov determinants, Eq. (\ref{rThroughOverlaps}), when combined with the explicit form of the overlaps through a determinants, Eqs.(\ref{OverlapsAsDdeterminant},\ref{NDefinition}),  makes use of the following the  determinants of four matrices $D(V,W;\epsilon^w),D({V},W, \epsilon^v)$, $N(V)$ and $N(W)$. Since these will be in heavy use in what follows, we shall label these matrices $D_{\pm}$, $N_V$, $N_W$, respectively, and define:
\begin{align}\label{DefinitionOfD1D2}
&D_{+\,\,\,a,b} = D_{a,b}(V,W;\epsilon^v), \quad \quad \quad N_{W\,\,a,b}=N_{a,b}(W)
\end{align}
where $a$ and $b$ run from $1$ to $P$\ for $l=1$ and from $1$ to $P-1$ for $l=-1$. Namely, for $l=-1$, we take only the upper $P-1\times P-1$ sub-matrix which does not include the rapidity close to $\varepsilon_m$. We likewise define $N_V =  N(V)$ and $D_{-} = D(W,V;\epsilon^w)\ $as a $P\times P$ matrices. The definitions made here are in anticipation for a representation of   $D_\pm$ in the  continuum limit as an operator acting on functions with certain analytic behavior, the determinants of which may in turn be computed using function theory methods. It is then the goal of this section to write expressions for  in terms of  $D_\pm$ and $N_V$, $N_W$.

In order to be able to proceed from  equation (\ref{rThroughOverlaps}) we need to compute the following matrix element $\<W, \epsilon^w | V\setminus \{v_\alpha\}\cup\{\varepsilon_m+\delta\}, {\epsilon}^w\>$. In principle this can be computed making use of (\ref{OverlapsAsDdeterminant}), which is general, but a more convenient form may be found, which reads:
\begin{align}\label{SlavnovIde}
&\nonumber\\
&\lim_{\delta\to0}\delta\frac{\<W, \epsilon^w | V\setminus \{v_\alpha\}\cup\{\varepsilon_m+\delta\}, {\epsilon}^w\>}{v_\alpha-\varepsilon_m}=2^P\det D_\alpha(W, V;\epsilon^w)
\end{align}
where
\begin{align}
D_\alpha(W,V;\epsilon^w)_{ab}=
\begin{cases}
D^{}_{ab}(W,V;\epsilon^w) &  b\neq \alpha\\
X_a (\varepsilon_m) & a=\alpha
\end{cases}\end{align}
and
\begin{align}\label{DefBigX}\quad  \left[\vec X  \right]_b=\frac{1}{2(v_b-\varepsilon_m)^2}-\frac{\delta H^{}(\varepsilon_m)}{2(v_b-\varepsilon_m)}
\end{align}
Cramer's rule may now be applied to obtain:
\begin{align}\label{MatrixElementsMainExpression1}
\<\mbox{in};l,p|c^\dagger_{m\sigma}|\mbox{in}\>^2 =\frac{\det \left[D(V,W;\epsilon^v)D(W, V;\epsilon^w)\right]}{\det \left[N(W)N(V)\right]}\left( 1-
\<\vec{X}|D^{-1}(V, W;\epsilon^v)|\vec 1\>\right),
\end{align}
where $|\vec 1\>$ signifies the vector with all elements equal to $1$. 

To obtain (\ref{SlavnovIde}) we take the Slavnov representation of the overlap, Eq. (\ref{Slavnov}), and rewrite $\det J$ as $\frac{\det C \Lambda J}{\det C \det \Lambda}$ , where $C_{ab} = \frac{1}{v_a-w_b}$ and $\Lambda_{ab}=\delta_{ab} \frac{\prod_c(w_a-w_c)}{\prod_{c\neq a}(w_a-w_c)}.$ Note that this is the same manipulation as the one done to obtain the representation (\ref{OverlapsAsDdeterminant}) from (\ref{Slavnov}). In fact, formally, these are also  the same $C$\ and $\Lambda $ matrices as those defined in (\ref{CauchyC}) and (\ref{LabmbdaMat})  (with the roles of $W$ and $V$ reversed). We stress,   however, that there  the set of rapidities $\{w_a\}_{a=1}^P$   is that set of rapidities featuring in the overlap, which  does not necessarily satisfy Richardson's equations, which for overlap (\ref{SlavnovIde})  would be  $V\setminus \{v_\alpha\}\cup\{\varepsilon_m+\delta\}$ and not simply $V$. As a result we obtain a different representation than the one in Eq. (\ref{OverlapsAsDdeterminant}), namely we arrive  Eq. (\ref{SlavnovIde}). The steps to obtain the result, however, are quite similar to the steps to obtain (\ref{OverlapsAsDdeterminant}) and so we do not repeat them here.

We start then from Eq. (\ref{MatrixElementsMainExpression1}) and write the result in terms of $D_\pm$, $N_V$ and $N_W.$  For $l=1$ the matrices $D( V, W;\epsilon^v),$ $D(W, V;\epsilon^w),$ $ N(V)$ and $N(W)$ are simply $D_{\pm}$, $N_V$, $N_W$ respectively, such that we may equivalently write:

\begin{align}\label{leq1DplusDminus}
&\<\mbox{in};1,p|c^\dagger_{m\sigma}|\mbox{in}\>^2= \frac{\det \left[D_+D_-\right]}{\det \left[N_V N_W\right]}\left[1-{\left\<\vec{X}\left|D_{-}^{-1}\right|\vec 1\right\>}\right]F_1,
\end{align}
where $|\vec 1\>$ denotes the vector with all elements equal to $1$. $F_1=1$ is introduced here since it allows us to generalize this result to the case where there are rapidities around $\varepsilon_m$. Indeed in Appendix \ref{MatrixDecompositionAppendix} it is shown that inclusion of such close rapidities only results in the modification of $F_1$, such that it remains a factor depending only on the configuration of rapidities around $\varepsilon_m$. We shall term such factor 'local factors' and encounter another such factor for the case $l=-1$.  Equation (\ref{leq1DplusDminus}) is in a form which is suitable for  computations in the continuum limit
 as $D_\pm$ will have a representation as operators in that limit, the coarse grained local factor may then be surmised by different means.

We wish, in the same spirit that lead to (\ref{leq1DplusDminus}), to write the result in Eq. (\ref{MatrixElementsMainExpression1}) in terms of $D_\pm, $ now for $l=-1$. To do so let us write more explicitly $D(V,W, \epsilon^v)$ in terms of  $D_\pm$, correct to leading order in $\iota$. We have:
\begin{align}
D(V,W;\epsilon^v) = \left( \begin{array}{cc} 
D_+ & \vec{Y} \\
\frac{\vec{Z}^t}{w_P-\varepsilon_m} & \frac{\delta H(\varepsilon_m)}{w_P-\varepsilon_m}
\end{array}  \right)  
\end{align}
where 
\begin{align}\label{DefYandZ}
\left[\vec{Z}\right]_b =\frac{1}{2} \frac{1}{w_b-\varepsilon_m},\quad  \left[\vec{Y}\right]_a =\frac{1}{2} \frac{1}{(w_a-\varepsilon_m)^2} .
\end{align}
For $N(W)$  we have (again to leading order):
\begin{align}
N(W) = \left( \begin{array}{cc} 
N_W & 0 \\
0 & \sum_{\epsilon_i \notin \epsilon^v }\frac{1}{(w_P-\varepsilon_i)^2}
\end{array}  \right).\end{align}  
It is a matter of expanding a determinant in minors to obtain the following result:
\begin{align}
&\det[D(V, W;\epsilon^v)] =\frac{\det D_{+}}{w_P-\varepsilon_m}\left(\delta H(\varepsilon_m)+\left\<\vec{Z}\left| D_{+}^{-1}\right|\vec{Y}\right\>\right).
\end{align}
In (\ref{MatrixElementsMainExpression1}) we also have the factor $1-{\left\<\vec{X}\left|D_-^{-1}\right|\vec 1\right\>}$. The vector $\<\vec X|$ can be seen to be given, in leading order, by  $-\frac{1}{w_P-\varepsilon_m}\<\vec{Z}|$, by noting that $\delta H(\varepsilon_m) = \frac{1}{w_P-\varepsilon_m}$, again to leading order. Thus we have to leading order $
 1-{\left\<\vec{X}\left|D_-^{-1}\right|\vec 1\right\>}=\frac{\left\<\vec{Z}\left|D_{-}^{-1}\right|\vec 1\right\>}{w_P-\varepsilon_m}.$
Substitution into (\ref{MatrixElementsMainExpression1}) to give:
\begin{align}\label{leqMinus1DplusDminus}
& \<\mbox{in};-1,p|c^\dagger_{m\sigma}|\mbox{in}\>^2= \frac{\det \left[D_+D_-\right]}{\det \left[N_VN_W\right]}\left(\delta H^{F}+\left\<\vec{Z}\left|D_{+}^{-1}\right|\vec{Y}\right\>\right)\left\<\vec{Z}\left|D_{-}^{-1}\right|1\right\>F_{-1},
\end{align}
where $F_{-1}=\frac{\iota}{(w_P-\varepsilon_m)^2h'^{(W)}(w_P)}$ is the 'local factor' when there is only one rapidity around $\varepsilon_m$ and we define  $\delta H^F = \delta H(\varepsilon_m)$.  Note that indeed $h'^{(W)}(w_P),  $ which appears in $F_{-1}$ depends to leading order only on the configuration of one particle levels around $\varepsilon_m$. 

In Appendix  \ref{MatrixDecompositionAppendix}  the case where there are additional close rapidities is treated, with the result is that (\ref{leqMinus1DplusDminus}) holds  $F_{-1}$ takes on a more complicated form, but remains a 'local factor, namely one that depends only on the local arrangement of rapidities and singly occupied levels around $\varepsilon_m$  and  $\delta H^F$ becomes the far field at $\varepsilon_m$:
\begin{align}
\label{deltaHF}
\delta H^F = \sum_{w_c\notin J} \frac{1}{\varepsilon_m-  w_c}  -  \sum_{v_c\notin J } \frac{1}{\varepsilon_m-  v_c}   ,
\end{align}
where $J$ is a given by $J=[\varepsilon_m - K\iota, \varepsilon_m + K\iota]$ and $K$ satisfies $1 \ll K \ll P$.  Namely, $\delta H^F$ is the contribution of charges far away from $\varepsilon_m$ to $\delta H(\varepsilon_m)$. 

We shall also need the near field version of $\delta H(\xi)$, which we denote by $\delta H^N(\xi)$, given by:
\begin{align}\label{deltaHN}
\delta H^N(w_b) = \sum_{c\neq b, w_c\in J} \frac{1}{w_b-w_c} - \sum_{c, w_c\in J} \frac{1}{w_b - v_c}.
\end{align}

The results of this section, Eqs. (\ref{leq1DplusDminus},\ref{leqMinus1DplusDminus}), gives the coherence factors in terms of a local factor times objects which have a representation in the continuum limit as operators, the latter facilitating the derivation of the explicit final result, Eq. (\ref{IntroductionResults}).   

\section{The Slavnov Matrix as an Operator   \label{ElectrostaticAnalogy}}

\subsection{Basic properties of the electric fields}
The fact that the Richardson equations have an electrostatic analogy has already been mentioned above. The electrostatic field and the electrostatic potential will feature heavily in what follows.
We discuss some basic properties of these fields. We shall be interested in the jump discontinuity and average across the arc of these fields. We thus define  $f^J(\xi)$ as the jump discontinuity of a function, $f,$ over the branch cut at point $\xi$:
\begin{align}
f^J(\xi) = f(\xi^+) - f(\xi^-).
\end{align}
where $\xi^+$ and $\xi^-$ are points slightly to the right and to the left of the arc respectively. Similarly we define the average value of the functio across the arc by $f^A$:
\begin{align}
f^A(\xi) = \frac{f(\xi^+) + f(\xi^-)}{2}.
\end{align}

The jump discontinuity of the function $h^{(V)}(\xi)$ over the arc is proportional to the line density of the charges there, $\sigma (\xi)$, measured in length units of $\iota$ as follows:
\begin{align}\label{discHisSigma} 
 h^{(V)J}(\xi)  =  2 \pi i \sigma(\xi),
\end{align}
where the number of rapidities in an interval $|d\xi|$ on the arc is equal to $\sigma(\xi) \frac{|d\xi|}{\iota}$.
We note that the average value of $h$ over the arc  is zero, which we denote by:
\begin{align}\label{avrghis0}
h^{(V)A}(\xi)  =  0.
\end{align}
This is derived from (\ref{hIntegral}).

We similarly $\delta H, $  defined in (\ref{DeltaHDef}) can be written as:
\begin{align}\label{deltahContinuum}
\delta H(\xi) =  H^{(W)}(\xi) - H^{(V)}(\xi) 
\end{align}
The property in Eq.  (\ref{avrghis0})  extends to $ \delta H$:
\begin{align}\label{deltahavrgis0}
 \delta H^A(\xi)  =  0,
\end{align}
since $\delta H$ is then truly a variation of $H$, and varying (\ref{hIntegral}) shows that the property (\ref{deltahavrgis0}) holds. The jump discontinuity of $\delta H$ is  equal to:
\begin{align}\label{discdeltaphi}
\delta H^J = 2 \pi i\partial_\xi\left[\sigma(\xi)\delta w(\xi)\right],
\end{align}
where $\delta w(\xi)$ is given by $\delta w(\xi)  = \frac{v_a-w_a}{\imath}$ for $v_a$ close to $\xi$ and $\partial_\xi$ is a derivative along the arc. To see this consider the following approximation for $\delta H$ far away from the arc:
\begin{align}
\delta H \simeq \partial_\xi \sum_a\frac{\iota\delta w_a}{\xi-w_a}.
\end{align}   
The sum is a function of $\xi$ which has a jump discontinuity of $2\pi i {\sigma(\xi)} \delta w (\xi),$ and so after the derivative is taken we have (\ref{discdeltaphi}). 

 \subsection{Opertor form of $D_\pm$}

Our purpose now is to compute the continuum limit of the matrices $D_\pm$. Consider then the action of $D_+$ on  $\vec{y}$ to produce $\vec{x}$:
\begin{align}\label{DonYisX}
D_{+} \vec y= \vec x
\end{align}

Let us parameterize the arcs by the arc length, $s$. For a point parameterized by $s$ on the arc, there exists some $i$ for which the rapidity, $w_i$, is the closest rapidity to this point. This defines a function $i(s)$. Conversely we define the function $s(i)$ to be the value the parameter $s$ takes on the point, $w_i$, on the arc.  We may represent $x_a$ by a continuous function, $x(s)$, of the parameter $s$, demanding:
\begin{align}\label{xOFsDefinition}
x(s(i)) = x_i,
\end{align}
while interpolating its value smoothly between the discrete set of point $s(i)$, for which (\ref{xOFsDefinition}) defines its value. A respective definition gives an interpolation $y(s)$ of the vector $y_i$.

Let us set a large number $K$. Define $r=K \iota$. We assume that $y(s)$ hardly changes on the scale $\iota$, which is the scale of the distance between the rapidities. This means that $y(s)$ and $x(s)$ do not change much on the scale $r$ (since $K$ is large but does not scale with $\frac{1}{\iota}$).  Making use of the explicit expression for the $D_+$ matrix, given in (\ref{OverlapsAsDdeterminant}),  we may write the following approximation for (\ref{DonYisX}):
\begin{align}\label{xOFsSumIntegral}
x(s) \simeq y(s) \sum_{|s(a) - s|<r} D_{+\,\,a , i(s)} + \int \limits_{|s(a) - s'| \geq r}y(s')  \left( \frac{1}{(w(s) - w(s'))^2}  - \frac{\Delta H^{}(w(s))}{(w(s) - w(s'))} \right)  \frac{ds' \sigma(s')}{\iota},
\end{align}
where $\sigma(s') \frac{ds'}{\iota}$ is the number of rapidities in the line element $ds'$ and $\Delta H$ is defined as follows:
\begin{align}
\Delta H = \Delta H^{W,\epsilon^v}.\label{DeltaHNoSuperscript}
\end{align}
Note that 
\begin{align}
- \Delta H^{V,\epsilon^w} =\Delta H^{W,\epsilon^v} = \Delta H = -\frac{1}{2} \frac{1}{\xi-\varepsilon_m}\label{DeltaHExpicit}.   
\end{align}

The first term in (\ref{xOFsSumIntegral}) has to be computed explicitly as a function of $r$ assuming that $K = \frac{r}{\iota}$ is very large. To demonstrate the method of calculation, while leaving details to the appendix \ref{KSumsAppendix}, we consider the term $D_{+\,\,i(s),i(s)}$, which appears in the sum. The expression for  $D_{+\,\,i(s),i(s)}$ is taken from the diagonal term in  (\ref{OverlapsAsDdeterminant}). To compute this we would have to write down an expression for $\delta H^{N}(w(s))$ (where $\delta H^N$ is defined in (\ref{deltaHN})). Consider that the distance between the rapidities in this range is almost constant and approximately equal to $\frac{\iota}{\sigma(s)}$. The rapidities, $v_i$, have to leading order,  the same spacing as that of the $w_i$'s, given by $\frac{\iota}{\sigma(s)}$, but they are shifted by an amount $\iota  \delta w(s)$.  Thus:
\begin{align}\label{coth}
\delta H^{N}(w(s)) & \simeq \frac{\sigma (s)}{\iota} \left( \sideset{}{'}\sum_{i=-K}^{K} \frac{1}{k} - \sum_{i=-K}^{K} \frac{1}{k+\sigma(s) \delta w(s)} \right)  = \\
&= \frac{\sigma (s) \pi }{\iota} \coth \left[\sigma(s) \delta w(s) \pi \right] \nonumber
\end{align}
The sums are taken only from $-K$ to $K$ since the contribution from  rapidities far away can be shown to be negligible. The reason for this is that these contribute an amount proportional to $ \delta h^{A} $, which, as was mentioned earlier, is zero.

A similar approach allows to write down an expression for $\sum_{|s(b) - s|<r} D_{+\,\,i(s),b}$. All the hyper-trigonometric expressions, such as in (\ref{coth}) cancel. A detailed description of this is given in Appendix \ref{KSumsAppendix}. One is left with:
\begin{align}\label{DactionLineIntegral}
 x(s) =  \frac{\pi^2\sigma(s)}{\iota} (\sigma (s) \delta  w(s))' y(s) + \partial_{w(s)} \dashint \frac{ \frac{\sigma(s')}{\iota} y(s')}{w(s') - w(s)} ds'  - \dashint \frac{\frac{ \sigma(s')}{\iota} \Delta H(s) y(s')}{w(s') - w(s)} ds'.
\end{align}
We may further simplify (\ref{DactionLineIntegral})  by defining a vector $\tilde{y}(\xi)$ as follows:
\begin{align}
\tilde{y}(\xi) = \sum \frac{y_b}{\xi - w_b}.
\end{align}
The jump discontinuity of $\tilde y(\xi), $ encodes the vector elements $y_b$. We thus have:
\begin{align}
 \tilde{y}^J(w(s)) =2\pi y_{i(s)} \frac{\sigma(s)}{\iota}.
\end{align}
This relation allows us  to convert the line integrals in (\ref{DactionLineIntegral}) into contour integrals around the arcs:
\begin{align}
 x(s) =  \frac{1}{2}(\sigma (s) \delta  w(s))'  \tilde{y}^J(w(s)) + \frac{1}{2\pi i}\left[  \partial_{\xi} \oint \frac{ \tilde{y}(\xi')}{\xi'- \xi} d\xi'  - \Delta H^{}(w(s))\oint \frac{    \tilde{y}(\xi')}{\xi' - \xi}d\xi'\right]^A(w(s)),
\end{align}
where $\Delta H^{}(\xi)$ is the analytic continuation of $\Delta H(s)$ away from the cut. Taking into account (\ref{deltahavrgis0}) and (\ref{discdeltaphi}), the whole right hand side can be expressed as:
\begin{align}\label{xofSasAvrg}
 x(s) =   \left[  \delta H(\xi) \cdot  \tilde{y}(\xi) +  \frac{\partial_{\xi}}{2 \pi i} \oint \frac{ \tilde{y}(\xi')}{\xi' - \xi} d\xi'  -  \frac{\Delta H^{}(\xi)}{2 \pi i} \oint \frac{   \tilde{y}(\xi')}{\xi' - \xi} \xi' \right]^A(w(s))
\end{align}
We now define a function $\hat{x}(\xi)$ such that $\hat{x}^A(w(s)) = x(s)$.    
We may this function  to be equal to the term in the square brackets in (\ref{xofSasAvrg}). Furthermore the  contour integrals in the latter equation can be taken explicitly giving us:
\begin{align}\label{DactionContinuous}
\hat{x}(\xi) = \left[ \delta H(\xi) -  \Delta H^{}(\xi)  \right]  \tilde{y}(\xi) + \tilde{y}'(\xi)
\end{align}
The vectors $\vec{x}$ and $\vec{y}$ are related through equation (\ref{DonYisX}), namely by the action of $D_+$. The action of $D_+$ is then represented in equation (\ref{DactionContinuous}) as a differential operator $\mathcal{D}_+$ which acts as $ \mathcal{D}_+\tilde{y}  (\xi)= \left[ \delta H(\xi) -  \Delta H^{}(\xi)  \right]  \tilde{y}(\xi) + \tilde{y}'(\xi).$   Define now an electrostatic potential, $\Phi$ by:
\begin{align}
\partial_\xi \Phi(\xi)=  \delta H^{} - \Delta H^{} \label{PhiDefinition}
\end{align}
We have:
\begin{align}
\mathcal{D}_+ = e^{-\Phi} \partial_\xi e^{\Phi} = (\partial_\xi \Phi) + \partial_\xi.
\end{align}
The same procedure may be applied to $\mathcal{D}_-$, with the only difference now that $\Phi$ switches sign:
\begin{align}\label{DthorughdPhi}
\mathcal{D}_\pm = e^{\mp\Phi} \partial_\xi e^{\pm\Phi}
\end{align}
The adjoint action can also be obtained, we spare also here  the details, and write the result:
\begin{align}\label{DdaggerthorughdPhi}
\mathcal{D}_{\pm}^{t}= e^{\mp\tilde{\Phi}} \partial_\xi e^{\pm\tilde{\Phi}} \mp \oint \frac{ \Delta H^{}(\xi')}{\xi' - \xi}d \xi'
\end{align}
where 
\begin{align} \label{PhiLDefinition}
\partial_\xi \tilde{\Phi}(\xi)=  \delta H^{} + \Delta H^{}. 
\end{align}
The continuum limit of the matrics $N_V$ and $N_W$ may be obtained by noting that these matrices have the same continuum limit obtained by substituting $\delta H=\Delta H=0$ and as such the right action of which are  represented by $\partial_\xi.$

\section{Explicit Expressions for Electrostatic Potentials for Two Arcs}
To compute $\Phi$ and ultimately the determinants of the various matrices we shall make use of concepts from the theory of algebraic Riemann surfaces already discussed here in the context of the Abel map.  Indeed, the Abel map allows one to go back and forth from a description through function of $\xi$ and a description of function of $u_\xi$. This may be facilitated by the introduction of the Weierstrass elliptic functions, which is a set of functions that respect the periodicity of the rectangle, which is in a sense complete. The first such function is the Weierstass $\wp$-function. This function has a double pole at the origin, and is doubly periodic:
\begin{align}
\wp(u) \simeq \frac{1}{u^2 },\quad\wp (u + \pi ) = \wp(u); \quad \wp (u + \tau) = \wp(u).
\end{align}
Another function is the Weierstrass $\zeta$ function. This function has a simple pole at the origin. This means that it cannot be doubly periodic, since this would imply that there is a function with a single pole on the Riemann surface, which is impossible, instead it is (additively) quasi-periodic:
\begin{align}
\zeta(u)\sim \frac{1}{u},\quad \zeta( u + \pi) = \zeta (u) + 2 \eta; \quad \zeta(u+ \tau) = \zeta(u) +  2 \eta'
\end{align}
where $\eta =  \zeta(\frac{\pi}{2}) $ and  $\eta' =  \zeta(\frac{ \tau}{2}) $.
The $\zeta$ function is simply minus the integral of $\wp$, $\zeta'(u) = -\wp(u)$. If one integrates again the $\zeta$ function one obtain the logarithm of the $\sigma$ function, namely $\frac{\sigma'(u)}{\sigma(u)} = \zeta(u)$. The function $\sigma(u)$ has a zero at the origin and is  (multiplicatively)\ quasi-periodic:
\begin{align}\label{SigmaProperty}
\sigma(u+\pi) = - \sigma(u) e^{2 \eta(u+ \frac{\pi}{2})}; \quad \sigma(u+ \tau) = - \sigma(u) e^{2 \eta'(u+ \frac{ \tau}{2})}
\end{align}
One may conveniently build all elliptic functions and differentials, namely functions and differentials respecting the periodicity of the rectangle, by using these building blocks $\sigma$, $\zeta$ and $\wp$, and their derivatives. For example consider the function $\xi(u)$ defined by the Able map, Eq. (\ref{AbelMap}). By definition, the function is elliptic with periods $\pi$ and $\tau$. The function has a pole on the upper sheet at $u_\infty$ and on the lower sheet at $-u_\infty$ with residues $\pm\frac{\pi}{2\omega},$ respectively, where \begin{align}
&\omega= \int_{\mu_1-i \Delta_1}^{\mu_1+i \Delta_1}\frac{1}{R_{4}(\xi')} d\xi. \label{omegaDef}
\end{align}
This means that  $\xi(u)$ is given by:
\begin{align}
&\xi(u)=\frac{\pi}{2\omega}\left(\zeta(u_\xi-u_\infty)-\zeta(u_\xi+u_\infty)\right)+\mbox{const},
\end{align}
where 
Moreover, if we take the derivative of the previous equation with respect to $u_\xi$ and compare it to the derivative of (\ref{AbelMap}) with respect to $\xi$, we find the following identity:
\begin{align}\label{R4wpIndentity}
&R_4(\xi)=\frac{4\omega^2}{\pi^2}\left(\wp(u_\xi-u_\infty)-\wp(u_\xi+u_\infty)\right)
\end{align}

We may now compute the different electrostatic objects in the case of two arcs. We shall need $\Phi $ as defined in (\ref{PhiDefinition})  and $\delta H^F$ as defined in (\ref{deltaHF}).

We now compute the differential $d_{\xi} \Phi = \Phi' d \xi$  in case $l=1$, i.e, the same number of rapidities in the vicinity of $\varepsilon_m$ in $V$ and $W$. In this case this differential has a pole on the upper sheet originating from $ \delta Hd \xi$ at $\varepsilon_m$ with residue $\frac{1}{2}$, which however cancels with a pole with opposite residue in $\Delta H^{} d\xi$. The same cancellation occurs for the pole at infinity on the upper sheet. The only poles of $d \Phi$ are then on the lower sheet at $\varepsilon_m$ and infinity, whose residues can be computed to be given by $\pm1$, respectively. We conclude:
\begin{align}\label{dxiPhiCaseI}
d_\xi \Phi = du_{\xi} \left[ \zeta\left(u_{\xi} +u_{\varepsilon_m}\right) - \zeta\left(u_{\xi} + u_{\infty}\right) +\frac{2\eta }{\pi  } (u_{\infty}  - u_{\varepsilon_m})  +2 i p  \right], \quad l=1.
\end{align}
Indeed the differential on the RHS has the right poles and residues, is periodic  and  its integral, $\Phi$,  picks up $2\pi i p$ going around the right arc counterclockwise.   The latter is the desired behavior since the definition of $\partial_\xi\Phi$, in Eq. (\ref{PhiDefinition}), and $\delta H$  in Eq. (\ref{BigdeltaHDef}) dictates that within such a cycle there are $N_1$ (the number of rapidities on the right arc in the $|{\rm in}\>$ state)  poles with residue $+1$ and  and  $N_1-p$ poles with residue $-1$.  For $l=-1$, the pole on the lower sheet at $\varepsilon_m$ cancels  and we obtain:
\begin{align}
d_\xi \Phi=du_\xi \left[\zeta(u_\xi - u_{\varepsilon_m}) - \zeta(u_\xi + u_{\infty}) +\frac{2\eta }{\pi  } (u_{\infty}  + u_{\varepsilon_m})+2 i p  \right], \quad l=-1,
\end{align}
where here the $\Phi$ picks up $2i\pi (p+1)$ going around the right arc counterclockwise. We can recast both results into one by writing:
\begin{align}
d_\xi \Phi= du_\xi \left[\zeta(u_\xi +l u_{\varepsilon_m}) - \zeta(u_\xi + u_{\infty}) +\frac{2\eta }{\pi  } (u_{\infty}  - l  u_{\varepsilon_m})+2 i p  \right],
\end{align}
Taking the integral we obtain an expression for $\Phi$ itself:
\begin{align}\label{PhiCaseI}
 \Phi(u) =\log \left[\frac{\sigma(u+ l u_{\varepsilon_m})}{\sigma( u+ u_{\infty})} \frac{\sigma(2 u_{\infty})}{\sigma( u_{\infty} + l u_{\varepsilon_m})}\right] + \left( \frac{2\eta }{\pi } (u_{\infty}  -l u_{\varepsilon_m})+2 i p\right) ( u- u_{\infty}).
\end{align}
As for $\tilde{\Phi}(u)$ defined in (\ref{PhiLDefinition}), it is simply given by $\tilde{\Phi}(u)$=$-\Phi(-u)$.

We shall also need $d_{\varepsilon_m} \Phi$, which is easily derived from (\ref{PhiCaseI}) to be given by:
\begin{align}
d_{\varepsilon_m} \Phi =l du_{\varepsilon_m}\left[  \zeta( u_\xi +lu_{\varepsilon_m})  - \zeta(u_{\infty}+l u_{\varepsilon_m} ) - l\frac{2\eta}{\pi}( u_\xi-u_{\infty} ) \right].
\end{align}

We now compute
$\delta H^{F}$, the far field part of $\delta H$ at $\varepsilon_m$, defined in (\ref{deltaHF}). Since $\Delta H$  is purely a near field  (according to (\ref{DeltaHExpicit}), it is given by $-\frac{1}{2}\frac{1}{\xi-\varepsilon_m}$) its contribution is irrelevant to the far field of $\partial_\xi \Phi$  at $\varepsilon_m$ and so the far field of  $\partial_\xi \Phi$ at $\varepsilon_m$ is given by $ \delta H^F$. For $l=1,$ the function   $\partial_\xi \Phi$ is  regular at $\varepsilon_m$, namely,  it is purely  far-field and  we conclude that, for this case, $\delta H^F = \partial_\xi \Phi.$  For $l=-1$, on the other hand, the field $\partial_\xi\Phi$ has a pole at $\varepsilon_m$ which must be removed to obtain $\delta H^F$, namely $\delta H^F = \lim_{\xi\to\varepsilon_m}\partial_\xi\Phi(\xi)+\frac{1}{\xi-\varepsilon_m}$. Making use of $\frac{1}{\xi-\varepsilon_m}d\xi = \left[\zeta(u_\xi-u_{\varepsilon_m})-\zeta(u_\xi-u_{\varepsilon_m}) \right]du_\xi$ and $\frac{du_\xi}{d\xi} =\frac{\pi}{2\omega} R_4(\xi)^{-1}$ one obtains:
\begin{align}\label{deltaHFExplicit}
\delta H^{F}=\frac{\pi l}{2\omega} \frac{\zeta(2u_{\varepsilon_m})-\zeta(u_{\varepsilon_m}+lu_\infty)+\frac{2\eta}{\pi}(lu_\infty-u_{\varepsilon_m})+2ilp}{R_4(\varepsilon_m)}
\end{align}

\section{Computation of Coherence Factors \label{ComputationofRatios}}
\subsection{Log Derivative of Determinants}
We wish to compute  $\det D_+D_-$ appearing in (\ref{leq1DplusDminus},\ref{leqMinus1DplusDminus}). For this purpose, we shall first compute $d_{\varepsilon_m} \log \det D_+D_-$ and then integrate in the exponent. We have:
\begin{align}
\label{logDetToCompute}d_{\varepsilon_m} \log \det D_{+}D_-= \tr \left(D_{+}^{-1} d_{\varepsilon_m}D_{+} +D_{-}^{-1} d_{\varepsilon_m}D_{-} \right).
\end{align}
The trace can be taken by considering the matrix as an operator acting on continuous functions. This is not a trivial statement, since the operator representation was derived assuming that $D_\pm$ act on vectors, $\vec{y}$ whose elements, $y_i$  smoothly vary as a function of the index, $i, $ while the trace is taken over all vectors, including those  which do not vary smoothly. Nevertheless, the rapidly varying part of the trace drops out, as  shown in Appendix \ref{CanTakeDetOfOperatorsAppendix}.

We proceed by writing Eq. (\ref{logDetToCompute}) through the operator $\mathcal{D}_{\pm}$ and its inverse $\mathcal{D}_\pm^{-1}$ given by:
\begin{align}\label{Dminus1Explicit}
\mathcal{D}_\pm^{-1}= e^{\mp\Phi} \int_\infty^\xi d\xi e^{\pm\Phi}.
\end{align}
The trace is effected by letting the operator being traced over act on $|i\>$, which has a representation through the function $\tilde{f}_i = \frac{1}{\xi - w_i}$,  contracting with the  linear form, $\<i|$, which has a representation as a linear operator given by $\<i|g\> =\frac{\tilde{g}^J(w_i)}{2\pi i \sigma(s(i))}  $ and summing over $i$, which is equivalent to integrating over $s$ with measure $\sigma(s(i))$. The result is:
\begin{align}
d_{\varepsilon_m} \log \det D_{+}=\frac{1}{2\pi i} \int  \left[ \mathcal{D}_{+}^{-1}(  d_{\varepsilon_m} \mathcal{D}_{+} ) \tilde{f}_{i(s)}  \right]^J\!\!\!\!(w(s))  ds.
\end{align}
For notational ease let us make the following symbolic substitutions $\tilde{f}_{i(s)} \longrightarrow |s\>$ and $f^J(w(s)) \longrightarrow \<s|f\>$. This leads to the following notation:
\begin{align}\label{dlogdetcontinu}
d_{\varepsilon_m} \log \det D_{+}= \int ds \<s| \mathcal{D}_{+}^{-1}(  d_{\varepsilon_m} \mathcal{D}_{+} ) |s\>   ds.
\end{align}
Since $d_{\varepsilon_m} \mathcal{D}_{+} = \partial_\xi d_{\varepsilon_m} \Phi$, as can easily be derived from (\ref{DthorughdPhi}), we have:
\begin{align}\label{DetExpression}
d_{\varepsilon_m} \log \det D_{+}= \int ds \<s| d_{\varepsilon_m}\Phi'(u_s) \mathcal{D}_{+}^{-1}  |s\>
\end{align}
To obtain the trace we need a representation of the standard basis in terms of functions on which $\mathcal{D}_{\pm}^{-1}$ may act. Consider then a function, $f_s,$ the average of which over opposite points on two sides of the arcs is proportional to a delta function. For two arcs we have
\begin{align}\label{fsDef2}
d\xi \hat{f}_s(\xi) = du_\xi\left\{\zeta(u-u_s)-\zeta(u+u_s)+\alpha(u_s)\left[\zeta(u-u_{\varepsilon_m})-\zeta(u+u_{\varepsilon_m})\right]+\beta(u_s)\right\}.
\end{align}
Note that if $u_{s'}$ is a point just to the left or to the right of any one of the arcs, then a point on the respectively opposite side is given by $-u_{s'}$, thus, indeed we have:
\begin{align}
d\xi \hat{f}^A_s(\xi(u_{s'})) = du_{s'} \delta(u_s-u_{s'}),
\end{align} 
for any $\alpha(s)$ and $\beta(s)$. 

We may now compute $\mathcal{D}_{\pm}^{-1}|s\>$ or the function corresponding to it, which we define as $e^{\mp\Phi(u)}g_\pm(u)$. Namely,\begin{align}\label{DefOfInverseOperators}
&\mathcal{D}_{\pm}^{-1}|s\> \to e^{\mp\Phi(u)}g_\pm(u) .
\end{align}
The logarithmic derivative of the determinants is given by:
\begin{align}\label{logDifferentialg}
& d_{\varepsilon_m} \log \det \left[ \mathcal{D}_+ \mathcal{D}_- \right] =\oint du_s d_{\varepsilon_m}\Phi'(u_s)\left[ e^{-\Phi(u_s)}g_+(u_s)-e^{\Phi(u_s)}g_-(u_s)\right].
\end{align}
The function  $g_\pm(u)$ is given explicitly by:
\begin{align}
&g_\pm(u) =\intop_{u_\infty}^{u}e^{\pm\Phi(u')}\left\{\zeta(u'-u_s)-\zeta(u'+u_s)+\alpha_\pm(u_s)\left[\zeta(u'-u_{\varepsilon_m})-\zeta(u'+u_{\varepsilon_m})\right]+\beta _\pm(u_s)\right\}du'
\end{align}
Where $\alpha_\pm(u_s)$ and $\beta_\pm(u_s)$ are functions of $u_s,$ fixed by the following conditions:

\begin{enumerate}
\item{  Periodicity for  tanslations  of $u$ by $\pi$ -- a requirement originating in the fact that on the upper sheet all functions must be well-defined, and a translation by $\pi$ is closed contour in the upper sheet. We thus have:
\begin{align}
& e^{\mp\Phi(0)}g_\pm(0)=e^{\mp\Phi(\pi)}g_\pm(\pi)
\end{align}}
\item{ The function $e^{-\Phi(u)}g_+(u)$ should have no pole at $\varepsilon_m$. This leads to the condition for for $l=-1$:
\begin{align}
&g_+(u_{\varepsilon_m})=0, \quad l=-1
\end{align}
}
\item{ The function  $e^{\Phi(u)}g_-(u)$ should be regular at $\varepsilon_m$. This leads to the following condition:
\begin{align}
&\underset{u\to u_{\varepsilon_m}}{\rm Res}\left[e^{-\Phi(u)}\left(\zeta(u-u_s)-\zeta(u+u_s)+\alpha_-(u_s)\left[\zeta(u-u_{\varepsilon_m})-\zeta(u+u_{\varepsilon_m})\right]+\beta_-(u_s)\right) \right]=0.
\end{align}}
\end{enumerate}
These analytical conditions uniquely fix $\alpha_\pm$ and $\beta_\pm.$ The solution is written as:
\begin{align}
&\nonumber\\
&\alpha_\pm(u_s)=-\delta_{l,-1}\frac{d_\pm A_{\pm}(u_s)-b_\pm B_\pm(u_s)}{a_\pm d_\pm-b_\pm c_\pm}, \quad
& \beta_\pm(u_s)=- \frac{ c_\pm A_{\pm}(u_s)\mp\delta _{l,-1}a_\pm B_\pm(u_s)}{ b_\pm c_\pm\mp\delta _{l,-1}a_\pm d_\pm }\label{alphabeta}
\end{align}

where the following  constants ($u_s$ independent) are defined:
\begin{align}\label{IntegralDefinition}
&\nonumber\\
&a_\pm=\intop_{0}^{\pi}e^{\pm\Phi(u')}\left[\zeta(u'-u_{\varepsilon_m} )-\zeta(u'+u_{\varepsilon_m})\right]du'\quad b_\pm =\intop_{0}^{\pi}e^{\pm\Phi(u')}du'\\
& c_+=\intop_{u_{\varepsilon_m}}^{u_\infty}e^{\Phi(u')}\left[\zeta(u'-u_{\varepsilon_m} )-\zeta(u'+u_{\varepsilon_m})\right]du'\quad d_+ =\intop_{u_{\varepsilon_m}}^{u_\infty}e^{\Phi(u')}du' \\
& c_- =\zeta(2u_{\varepsilon_m})-\zeta(u_\infty+u_{\varepsilon_m})+(u_\infty+u_{\varepsilon_m})\frac{2\eta}{\pi}+2 i p \quad d_{-}=1
\end{align}
along with  the following functions of $u_s$:
\begin{align}
&A_\pm(u_s)=\int_0^{\pi} e^{\pm\Phi(u')}\left[\zeta(u'-u_s )-\zeta(u'+u_s)\right]du' \label{ADef}\\
&B_+(u_s)=\int_{u_{\varepsilon_m}}^{u_\infty} e^{\Phi(u')}\left[\zeta(u'-u_s )-\zeta(u'+u_s)\right]du',\quad B_-(u_s)=\zeta(u_{\varepsilon_m}-u_s)-\zeta(u_{\varepsilon_m}+u_s) \nonumber
\end{align}

\begin{figure}[h!!!]
\includegraphics[width=6cm]{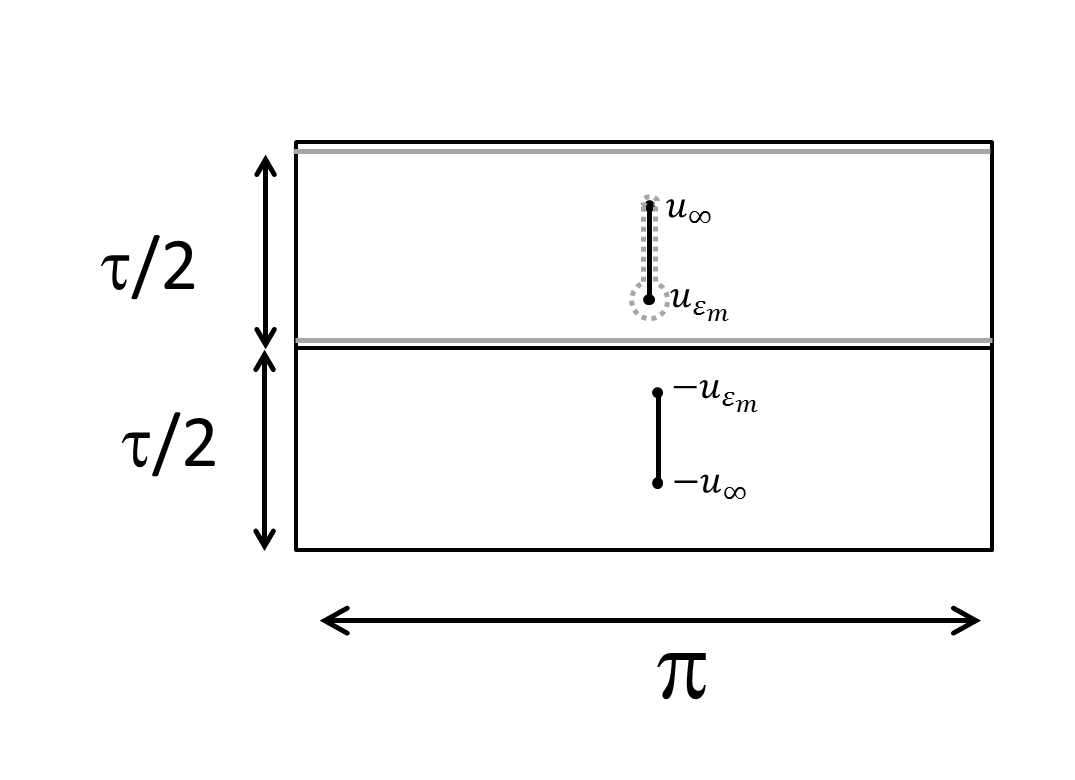}
\caption{Path of contour integration for $u_s$. The branch cuts are drawn in heavy black lines. Original path is drawn in heavy gray line, while the formed line is drawn in  dashed gray line. The path is deformed as to encircle the branch cut between $u_\infty$ and $u_{\varepsilon_m}$. The deformed path contains a small circle of radius $r$ and a part that tightly winds around the branch cu. \label{usContour}}
\end{figure}

We must substitute $g_\pm(u)$ into (\ref{logDifferentialg}) and compute the integral. We first separate the logarithmic divergence at $u\to u_s$ in  $g_{\pm}(u)$ by  splitting $g_\pm(u)$ into two parts, $g_\pm(u) = g^{(1)}_\pm(u)+g^{(2)}_\pm(u)$ as follows:
\begin{align}
g^{(1)}_\pm(u)&=e^{\pm\Phi(u_s)}\intop_{u_\infty}^{u}\left[\zeta(u'-u_s)-\zeta(u'+u_s)\right] du' =e^{\pm\Phi(u_s)}\log\left(\frac{\sigma(u-u_s)\sigma(u_\infty+u_s)}{\sigma(u+u_s)\sigma(u_\infty-u_s)}\right)\\
g^{(2)}_\pm(u) &=\intop_{u_\infty}^{u}\left(e^{\pm\Phi(u')}-e^{\pm\Phi(u_s)}\right)\left(\zeta(u'-u_s)-\zeta(u'+u_s)\right)+ \nonumber\\
 &+\int_{u_\infty}^u e^{\pm\Phi(u')}\left[\alpha_\pm(u_s)\left(\zeta(u'-u_{\varepsilon_m})-\zeta(u'+u_{\varepsilon_m})\right)+\beta _\pm(u_s)\right] du'\label{g2Def}
\end{align}
As a result $g_\pm^{(1)}$ contains the logarithmic divergence at $u\to u_s$  while $g_{\pm}^{(2)}$  is regular at this point. The terms proportional to $g_{\pm}^{(1)}$ cancel in  (\ref{logDifferentialg}) and the  calculation is reduced to computing the contribution of $g_\pm^{(2)}$ to (\ref{logDifferentialg}). Here one must substitute
(\ref{alphabeta}) into (\ref{g2Def}) and compute:
\begin{align}
\oint  du_s d_{\varepsilon_m}\Phi'(u_s)e^{\mp \Phi(u_s)} g^{(2)}_\pm(u_s).\label{g2logDet}
\end{align}
Since $\alpha_\pm(u_s)$ and $\beta_\pm(u_s)$ are linear combinations of $A_\pm(u_s), B_\pm(u_s)$ and $C_\pm(u_s),$ the result may be obtained by computing the integrals  in (\ref{g2logDet}) that one obtains   by substituting $A_\pm(u_s), B_\pm(u_s)$ and $C_\pm(u_s) $ for $\alpha_\pm(u_s)$ and $\beta_\pm(u_s)$, respectively, and taking the appropriate linear combinations of the results. The computations are  straightforward but lengthy. We shall give one example of such an integral and then cite the final result for (\ref{logDifferentialg}), rather than writing the full derivation, which is rather mechanical. We consider, then, as an example, the integral involving $B_+(u_s)$. First we write $B_+(u_s)$ as:
\begin{align}
B_+(u_s)=e^{\Phi(u_s)}\log\left(\frac{\sigma(u_s-u_{\varepsilon_m})}{\sigma(u_s-u_\infty)}\frac{\sigma(u_s+u_\infty)}{\sigma(u_s+u_{\varepsilon_m})}\right)+\int_{u_{\varepsilon_m}}^{u_\infty} (e^{\Phi(u')}-e^{\Phi(u_s)})\left[\zeta(u'-u_s )-\zeta(u'+u_s)\right]du.
\end{align}
We shall calculate the contribution  from the first term on the right hand side of this equation to  (\ref{logDifferentialg}). We denote the integral, which appears only for the case $l=-1$,  by $I$: \ 
\begin{align}\label{CExpression}
I=\oint  du_s d_{\varepsilon_m}\Phi'(u_s)\intop_{u_\infty}^{u_s}du'e^{\Phi(u')} \log\left(\frac{\sigma(u_s-u_\infty)}{\sigma(u_s-u_{\varepsilon_m})}\frac{\sigma(u_s+u_{\varepsilon_m})}{\sigma(u_s+u_\infty)}\right)
\end{align}
The integral over $u_s$ may be performed first and the integral over $u'$ second. We note that, as a function of $u_s$, the integrand has  branch cuts between $u_{\varepsilon_m}$ and $u_\infty$ and between $-u_{\varepsilon_m}$ and $-u_\infty,$ where the integrand has a jump discontinuity of $2\pi i  e^{\Phi(u') d_{\varepsilon_m}\Phi'(u_s)}$. The contour integral over $u_s$ is deformed, as shown in  Fig. \ref{usContour},  to encircle the branch cut between $u_{\varepsilon_m}$ and $u_\infty$ in a contour which contains an almost closed small  circle of radius $r$ around $u_{\varepsilon_m}$, while the rest of the contour winds tightly  around the branch cut. Such a separation is necessary due to the singular behavior around $\varepsilon_m$.  Computing separately the contribution of the circular part, which in turn is computed by performing an expansion of  the integrand around $u_{\varepsilon_m},$ and the contribution of the linear part, in which the simple form of the jump discontinuity  of the integrand (namely $2\pi i  e^{\Phi(u') d_{\varepsilon_m}\Phi'(u_s)}$)  is made use of,  one obtains:    
\begin{align}
I=-du_{\varepsilon_m}\left[\frac{1}{r}+\zeta(u_{\varepsilon_m}+u_\infty)-\zeta(u_{\varepsilon_m}-u_\infty)-\zeta(2u_{\varepsilon_m})\right] d_++\int_{u_\infty}^{u_{\varepsilon_m}-r} du_s d_{\varepsilon_m} \Phi'(u_s)\int_{u_\infty}^{u_s} du'e^{\Phi( u')},  
\end{align}
  where only the first terms in an expansion in small $r$ are retained. The integral here may be further simplified by integration by parts over $u_s$  and making use of the explicit form of $\Phi$, Eq. (\ref{PhiCaseI}).  The result is:
\begin{align}
I=-du_{\varepsilon_m}\left(\zeta(u_{\varepsilon_m}+u_\infty)-(u_{\varepsilon_m}-u_\infty)\frac{\eta}{\omega}-\zeta(2u_{\varepsilon_m})\right)d_+-\intop_{u_\infty}^{u_{\varepsilon_m}}du'd_{\varepsilon_m}e^{\Phi(u')}.
\end{align}

Continuing the calculation by applying similar methods to all the other integrals encountered in (\ref{g2logDet}), wherein the calculation displays many cancellations of the complicated expressions, one obtains:
\begin{align}\label{DerivativeFinal}
d_{\varepsilon} \log \det \left[ \mathcal{D}_+ \mathcal{D}_- \right] &= d_{\varepsilon_m}\log\left[\frac{(b_+c_+-a_+d_+\delta_{l,-1})(b_-c_- +a_-d_-\delta_{l,-1})}{\left(c_+c_-\right)^{\delta_{l,1}}}\right]+ \nonumber\\ &+\left(\zeta(u_{\varepsilon_m}+u_\infty)+\zeta(u_{\varepsilon_m}-u_\infty)-\frac{4u_{\varepsilon_m}\eta}{\pi}-2 i p\right)\delta_{l,-1,}
\end{align}
which  upon integration yields
\begin{align}\label{LogDetFinal}
\det \left[\frac{{      D}_{+} {D}_{-}}{N_V N_W}\right]&=\tilde C_{p,l}\frac{\left(b_+c_+-a_+d_+\delta_{l,-1}\right)(b_-c_- +a_-d_{-}\delta_{l,-1})}{\left(c_+c_-\right)^{\delta_{l,1}}}   \times \nonumber \\ &\times\left(\frac{e^{\left(u_\infty-u_{\varepsilon_m}\right)\left((u_\infty+u_{\varepsilon_m})\frac{2\eta}{\pi}+2 i p\right)}\sigma\left(u_\infty-u_{\varepsilon_m}\right)\sigma\left(u_\infty+u_{\varepsilon_m}\right)}{\sigma\left(2u_\infty\right)}\right)^{\delta_{l,-1}}
\end{align}
where $\tilde C_{p,l}$ are the integration constants to be determined later.

\subsection{Matrix Elements }

We compute the different matrix elements appearing in (\ref{leq1DplusDminus}) and (\ref{leqMinus1DplusDminus}). First, we compute $\<\vec{X}|D_-^{-1}|1\>,$ in which case we specialize to $l=1$ as this matrix element appears only in the expression for the matrix element for the case $l=1$,  Eq. (\ref{leq1DplusDminus}). The vector $\vec {X}$ is given in (\ref{DefBigX}), and it may be  represented by a function $f_{\vec{X}}(\xi) = \frac{1}{2(\xi - \varepsilon_m)^2} -\frac{\delta H(\varepsilon_m)}{2(\xi - \varepsilon_m)}$, whose average value over the arcs is given by $\vec{X}$. Applying $D_-^{t-1}$ on $f_{\vec{X}}$ gives:
\begin{align}\label{InnerIntu}
\mathcal{D}_-^{t -1}f_{\vec{X}}(\xi)=\frac{\frac{e^{2\left(u_{\varepsilon_{m}}-u_\xi\right)\left(\left(u_{\varepsilon_{m}}-u_{\infty}\right)\frac{\eta}{\pi}- i p\right)}}{\sigma(u_{\infty}-u_{\varepsilon_{m}})}\frac{\sigma\left(u_\xi-u_{\infty}\right)}{\sigma\left(u_\xi-u_{\varepsilon_{m}}\right)}
+\zeta(u_{\infty}+u_{\varepsilon_m})-\zeta(u_{\infty}-u_{\varepsilon_m})+\zeta(u_\xi-u_{\varepsilon_m})-\zeta(u_\xi+u_{\varepsilon_m})}{\frac{2\omega}{\pi}\left(\wp(u_
{\varepsilon_m}+u_\infty)-\wp(u_
{\varepsilon_m}-u_\infty)\right)}
\end{align}
To compute $\<\vec{X}|D_-^{-1}|1\>$  one should sum up over all vector elements of $ \<\vec{X}|D_-^{-1}$, which is equivalent to taking the contour integral of $\mathcal{D}_-^{\dagger -1}f_{\vec{X}}(\xi)$ around the arcs,  $\<\vec{X}|D_-^{-1}|1\>=\oint \mathcal{D}_-^{t -1}f_{\vec{X}}(\xi) d\xi$, an integral which can be performed by expanding the contour to infinity and picking up poles. The result is:
\begin{align}
\label{leq1MatElemContinuum}\<\vec{X}|D_{-}^{-1}|1\>=1-\frac{e^{2\left(u_{\infty}-u_{\varepsilon_{m}}\right)\left(\frac{\eta}{\pi}\left(u_{\infty}-u_{\varepsilon_{m}}\right)+{i p}\right)}}{\left(\wp(u_
{\varepsilon_m}-u_\infty)-\wp(u_
{\varepsilon_m}+u_\infty)\right)\sigma\left(u_{\infty}-u_{\varepsilon_{m}}\right)^2}
\end{align}

We now turn to evaluation of the matrix elements appearing in Eq. (\ref{leqMinus1DplusDminus}), namely for  the case $l=-1$. The matrix elements in question are $\<\vec{Z}|D_\pm^{-1}|\vec{Y}\>$.  We first compute $\<\vec Z|D_\pm^{-1}$, making use of $g_{\vec{Z}}(\xi) = \frac{1}{2(\xi - \varepsilon_M)}$, the result is:
\begin{align}\label{Guess1}
&\mathcal{D}_+^{t -1} g_{\vec{Z}}(\xi)=\frac{d_+\intop_{0}^{\pi}e^{\Phi(u')}\left[\zeta(u'+u_\xi)-\zeta(u'+u_\infty)\right]du'-b_+\intop_{u_{\varepsilon_m}}^{u_\infty}e^{\Phi(u')}\left[\zeta(u'+u_\xi)-\zeta(u'+u_\infty)\right]du'}{b_+c_+-a_+d_+}\\
&\mathcal{D}_-^{t -1}g_{\vec{Z}}(\xi)=-\frac{e^{2\left(u_\xi-u_{\varepsilon_m}\right)\left((u_\infty+u_{\varepsilon_m})\frac{\eta}{\pi}+ i p\right)}\sigma(2u_{\varepsilon_m})\sigma(u_{\infty}-u_\xi)}{\sigma(u_\infty-u_{\varepsilon_m})\sigma(u_\xi+u_{\varepsilon_m})}\label{Guess2}
\end{align}

To compute $\<\vec Z|D_-^{-1}|1\>$ one takes a contour integral of $\mathcal{D}_-^{t -1}g_{\vec{Z}}(\xi)$ around the arcs to obtain:
  \begin{align}
\label{MatElemleqMinus1Dminus}\<\vec Z|D_-^{-1}|1\>=-\frac{e^{2\left(u_\infty-u_{\varepsilon_m}\right)\left((u_\infty+u_{\varepsilon_m})\frac{\eta}{\pi}+ i p\right)}\sigma(2u_{\varepsilon_m})}{\sigma(u_{\varepsilon_m}-u_\infty)\sigma(u_{\varepsilon_m}+u_{\infty})}\frac{\pi}{2\omega}
\end{align}
To compute $\<\vec Z|D_+^{-1}|\vec{Y}\>$ one makes use of the  $\<\vec Z|D_+^{-1}|\vec{Y}\> =\oint \frac{1}{2(\xi-\varepsilon_m)^2}\mathcal{D}_+^{t -1} g_{\vec{Z}}(\xi) $. Indeed, multiplying $\mathcal{D}_+^{t-1} g_{\vec{Z}}(\xi)$ by $\frac{1}{2(\xi-\varepsilon_m)^2}$ just multiplies the jump discontinuity of  $\mathcal{D}_+^{t -1} g_{\vec{Z}}(\xi)$ by this factor and the contour integral then sums up over all vector elements. Making use of (\ref{deltaHFExplicit}), the result may be written as:\begin{align}
&\label{MatElemleqMinus1DPlus}\left\<\vec{Z}\left|D_{+}^{-1}\right|\vec{Y}\right\>=\frac{\pi}{R_4(\varepsilon_m)4\omega}\frac{d_+\Gamma-b_+\Lambda}{a_+d_+-b_+c_+-}-\delta H^F
\end{align}
where:
\begin{align}\label{GammaDef}
&\Gamma=\intop_{0}^{\pi}e^{\Phi(u')}Q(u')du' 
\end{align}
and
\begin{align}
&\Lambda=\intop_{u_{\varepsilon_m}}^{u_\infty}e^{\Phi(u')}Q(u')du'\label{LambdaDef} \end{align}
where
\begin{align}
\label{QDef}Q(u)=\wp(u+u_{\varepsilon_m})+\frac{\pi}{2\omega}
 \delta H^F\cdot \left(\wp(u_{\varepsilon_m}-u_\infty)-\wp(u_{\varepsilon_m}+u_\infty)\right)\left(\zeta(u-u_{\varepsilon_m})-\zeta(u+u_{\varepsilon_m})\right)
\end{align}

\subsection{Putting The Result Together}
We can now compute the coherence factors by combining for $l=1$ Eqs. (\ref{leq1DplusDminus}), (\ref{LogDetFinal}) and (\ref{leq1MatElemContinuum}):
\begin{align}\label{ResultCase1}
\<\mbox{in};1,p|c^\dagger_{m\sigma}|\mbox{in}\>^2
=\frac{\pi^2 C_{p,1}}{\omega^2}
\frac{b_+b_-e^{2\left(u_{\infty}-u_{\varepsilon_{m}}\right)\left(\frac{\eta}{\pi}\left(u_{\infty}-u_{\varepsilon_{m}}\right)+ i p\right)}}{R_4\left(\varepsilon_m\right)\sigma\left(u_{\infty}-u_{\varepsilon_{m}}\right)^2}F_{1},
\end{align}
where  we have introduced  constants  $C_{p,l}$, which play a similar role as $\tilde C_{p,l}$.
For $l=-1$ we must combine  (\ref{leqMinus1DplusDminus}), (\ref{LogDetFinal}), (\ref{MatElemleqMinus1Dminus}) and (\ref{MatElemleqMinus1DPlus})\begin{align}\label{ResultCase2}
\<\mbox{in};-1,p|c^\dagger_{m\sigma}|\mbox{in}\>^2&=\frac{\pi^2 C_{p,-1}}{\omega^2}\left\{b_-c_- +a_-d_-\right\}\left\{d_+\Gamma-b_+\Lambda\right\}  \frac{e^{4\left(u_\infty-u_{\varepsilon_m}\right)\left((u_\infty+u_{\varepsilon_m})\frac{\eta}{\pi}+ i p\right)}\sigma\left(2u_{\varepsilon_m}\right)}{\sigma\left(2u_\infty\right)R_4\left(\varepsilon_m\right)}F_{-1.}
\end{align}

We would now like to further simplify the results obtained for the matrix elements (\ref{ResultCase1},\ref{ResultCase2}) by performing an explicit calculation of the following quantities:
\begin{align}\label{Quan}
b_\pm,\quad b_-c_- +a_-d_-,\quad \Gamma
\end{align}

\begin{figure}[h!!!]
\includegraphics[width=3cm]{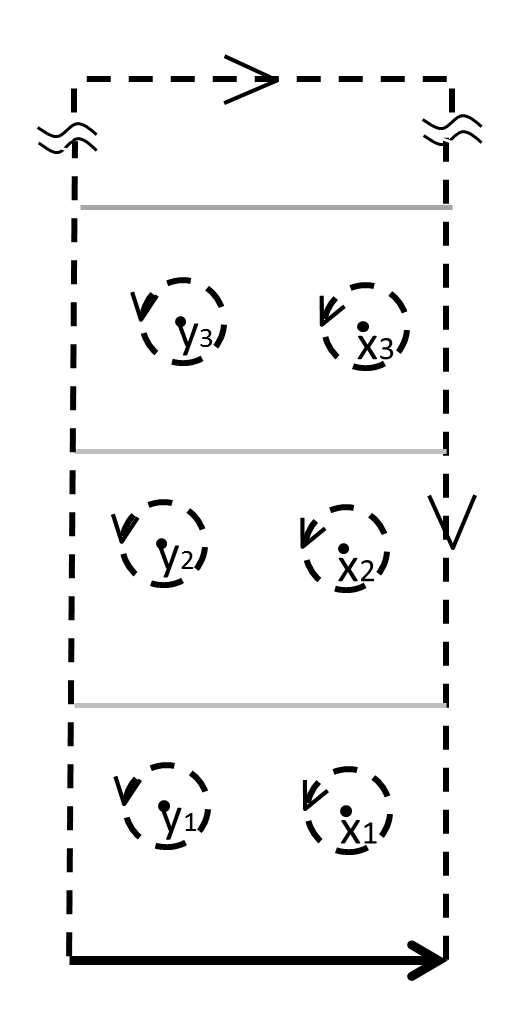}
\caption{Path of contour integration for  $\pm{\rm Im}\left[ u_{\varepsilon_m}-l u_\infty+\tau p\right] >0$. The original path from $0$ to $ \pi$ is drawn in solid line, while the deformed path is drawn in dashed line. The gray lines are placed at intervals of $\tau$ consistent with the quasi-periodicity of the integrand.   The integration over the left and right vertical dashed lined cancel due to periodicity of the integrand. The integrand tends to zero on the far top horizontal portion of the dashed line such that this part of the deformed integration path  may be omitted, as well. The only contribution left is that from the integration paths around the poles at $x_n$ and $y_n$ that lie between the original contour of integration and the deformed one. \label{IntPath}}
\end{figure}

Consider first $b_\pm = \int_0^{\pi}e^{\pm \Phi(u')} du'$, as defined in (\ref{IntegralDefinition}). For $\pm{\rm Im}\left[ u_{\varepsilon_m}-l u_\infty+\tau p\right] >0$, we now deform the  integration path in this integral  to a rectangular path plus paths that surround the poles of $ e^{\pm\Phi}$, as shown in  the  in Fig. \ref{IntPath} (for   $\pm{\rm Im}\left[ u_{\varepsilon_m}-l u_\infty+\tau p\right] <0$ the integration path is the mirror image over the real axis of the one shown in Fig. \ref{IntPath}). The vertical segments of the path cancel due to the periodicity of the integrand for translation in $\pi$. The horizontal piece tends to zero  if the rectangle is high enough since the integrand tends to zero for large and positive  imaginary $u'$  (or large and negative imaginary $u'$  for $\pm{\rm Im}\left[ u_{\varepsilon_m}-l u_\infty+\tau p\right] <0$ ).     The latter statement being due to the  quasi-periodicity of $e^\Phi$:
\begin{align}\label{vanishing}
e^{\pm\Phi(u+\tau)}=e^{\pm2  i(u_{\varepsilon_m}-l u_\infty+\tau p)}e^{\pm\Phi(u)},
\end{align}
which is a consequence of the definition of $\Phi$, Eq. (\ref{PhiCaseI}), and the quasi-periodicity of the Weierstrass $\sigma$-function, as given in  Eq. (\ref{SigmaProperty}). The integral is thus reduced only to the contribution of the  poles of the integrand, $e^{\pm \Phi}$, which are related to the zeros, $u_{i,j}$ of the sigma function $\sigma(u), $ which in turn are known to form a lattice $u_{i,j} =  i\pi + j \tau $.  The poles which are then picked up by deforming the contour of integration are $x_n=-u_\infty+\pi+\tau n$ and $y_n=-lu_{\varepsilon_m}+\pi\delta_{l,1}+\tau(n-\delta_{l,-1})$. The computation of residues at this points is straight forward, we find:
\begin{align}\label{Resb}
&\underset{u\rightarrow y_n}{\rm Res}\left(e^{\Phi(u)}\right)=\gamma_+ e^{\kappa n};\quad \underset{u\rightarrow y_n}{\rm Res}\left(e^{-\Phi(u)}\right)=\gamma_- e^{-\kappa n}
\end{align}
where
\begin{align}
&\gamma_+=-\frac{\sigma\left(u_\infty- l u_{\varepsilon_m}\right)\sigma(2u_{\infty})}{\sigma(u_{\infty}+l u_{\varepsilon_m})} e^{-4 u_\infty\left[(u_\infty-l u_{\varepsilon_m})\frac{\eta}{\pi}+ i p\right] }\nonumber\\
&\gamma_- = \frac{\sigma\left(u_\infty- l u_{\varepsilon_m}\right)\sigma(u_\infty+lu_{\varepsilon_m})}{\sigma(2u_{\infty})}  e^{2(u_\infty+l u_{\varepsilon_m}) \left[(u_\infty-l u_{\varepsilon_m})\frac{\eta}{\pi}+ i p\right] }\nonumber\\
&\kappa = 2 i (\tau p-u_\infty+l u_{\varepsilon_m})
\end{align}
The summation over the poles is then just a summation of  simple geometric series,
yielding :
\begin{align}\label{b+b-}
&b_\pm=\pm\frac{\gamma_\pm}{\left(1-e^{\mp\kappa}\right)}.
\end{align}
Substitution of these results into (\ref{ResultCase1})  and use of the  identity $\pi\eta'-\tau\eta=\pi i$, yields:
\begin{align}\label{ResultCase1-2}
\<\mbox{in};1,p|c^\dagger_{m\sigma}|\mbox{in}\>^2=\frac{\pi^2 C^1_p\sin^{-2}\left[u_{\varepsilon_m}-u_\infty+p \tau\right]}{4\omega^2R_4\left(\varepsilon_m\right)} {F}_1
\end{align}
The procedure just outlined for the computation of $b_\pm$ may be easily adapted to treat   $b_-c_- +a_-d_- $ and $\Gamma$. We  only cite the result:
\begin{align}
b_-c_- +a_-d_-&=\frac{\sigma(u_\infty-u_{\varepsilon_m})^2\exp\left[\frac{2}{\pi}(u_\infty+u_{\varepsilon_m})\left(\pi i p +(u_\infty+u_{\varepsilon_m})\eta\right)\right]}{\sigma(2u_\infty)\sigma(2u_{\varepsilon_m})\left(1-\exp\left[-2 i(u_\infty+u_{\varepsilon_m}-\tau p)\right]\right)}\\
  \Gamma&=-\frac{\sigma(u_\infty+u_{\varepsilon_m})\sigma(2u_\infty)\exp\left[-\frac{4u_\infty}{\pi}\left(\pi i p +(u_\infty+u_{\varepsilon_m})\eta\right)\right]}{\sigma(u_\infty-u_{\varepsilon_m})\left(1-\exp\left[2 i(u_\infty+u_{\varepsilon_m}-\tau p)\right]\right)}Q \nonumber(-u_\infty).
\end{align}
Substitution of these  expressions and those found for $b_+,b_-$ (\ref{b+b-}) into (\ref{ResultCase2}), yields
\begin{align}\label{ResultCase2-1}
&\<\mbox{in};-1,p|c^\dagger_{m\sigma}|\mbox{in}\>^2=\frac{\pi^2 C^{-1}_p\sin^{-2}\left[u_{\varepsilon_m}+u_\infty- p \tau\right]}{4\omega^2R_4\left(\varepsilon_m\right)}e^{2\left(u_\infty-u_{\varepsilon_m}\right)\left((u_\infty+u_{\varepsilon_m})\frac{\eta}{\pi}+ i p\right)}\Theta F_{-1}
\end{align}
where
\begin{align}\label{DefTheta}
&\Theta=  \frac{\sigma(u_\infty-u_{\varepsilon_m})\sigma(u_\infty+u_{\varepsilon_m})}{\sigma(2u_\infty)} (d_+Q (-u_\infty)- \Lambda)
\end{align}
We further simplify the result by showing that  $\Theta$ is given by a much shorter expression. To do so we first note that substituting this the definition of $\Theta$ the integral expressions for  $d_+, \Lambda,  $ Eqs. (\ref{IntegralDefinition},\ref{LambdaDef}),
and making use of the definition of $Q$, Eq. (\ref{QDef}), one obtains:
\begin{align}
\Theta \label{ThetaSimpleIntegra}=\intop^{u_\infty}_{u_{\varepsilon_m}}\frac{\sigma(u-u_{\varepsilon_m})}{\sigma(u+u_\infty)}\left(Q(-u_\infty)-Q(u)\right) du. 
\end{align}
 We make use  the following identity:
\begin{align}
&\frac{\sigma(u-u_{\varepsilon_m})}{\sigma(u+u_\infty)}\left(Q(-u_\infty)-Q(u)\right)\nonumber\\
&=\frac{\sigma(2u_{\varepsilon_m})\sigma(u-u_\infty)}{\sigma(u+u_{\varepsilon_m})\sigma(u_\infty-u_{\varepsilon_m})\sigma(u_\infty+u_{\varepsilon_m})}\left(\zeta(u-u_\infty)-\zeta(u+u_{\varepsilon_m})+2 i p+(u_\infty+u_{\varepsilon_m})\frac{2\eta}{\pi}\right)
\end{align}
which verified by comparing the poles and residues  of both sides of the equation. The integral over the latter expression may be taken to obtain:
\begin{align}
\Theta&=-e^{-2\left(u_\infty-u_{\varepsilon_m}\right)\left((u_\infty+u_{\varepsilon_m})\frac{\eta}{\pi}+ i p\right)}
\end{align}
Substituting this into (\ref{ResultCase2-1}) The matrix element for $l=-1$ then takes on a simpler form:
\begin{align}\label{ResultCase2-2}
\<\mbox{in};-1,p|c^\dagger_{m\sigma}|\mbox{in}\>^2=-\frac{\pi^2 C_{p,-1}\sin^{-2}\left[u_{\varepsilon_m}+u_\infty- p\tau \right]}{4\omega ^2R_4\left(\varepsilon_m\right)} F_{-1}
\end{align}

We are interested in  coarse grained quantities, $\overline{\<\mbox{in};l,p|c^\dagger_{m\sigma}|\mbox{in}\>^2} $, where the overline denotes coarse graining, namely an average over all possible $|{\rm in}\>$-states and a sum over all possible  $|{\rm out}\>$ states with the same coarse grained densities and the same $l$ and $p$. Taking into account (\ref{ResultCase1-2}) and (\ref{ResultCase2-2}), we have: 
\begin{align}\label{almostFinal}
&\overline{\<\mbox{in};l,p|c^\dagger_{m\sigma}|\mbox{in}\>^2}=\frac{l \pi^2 \sin^{-2}\left[u_{\varepsilon_m}-lu_\infty+lp \tau\right]}{4\omega^2 R_4\left(\varepsilon_m\right)} \overline{F_l}C_{p,l},
\end{align}
 where coarse graining is only over the local factor $F_l$ since only it depends on the microscopic arrangement of rapidities. Note that in (\ref{almostFinal}) we do not assume that the fermionic creation operator acts on level which is not singly occupied. Indeed, the results of section \ref{OneMatrixElemSection} show that the same form, given in Eqs.  (\ref{ResultCase1-2},\ref{ResultCase2-2}), is obtained whether the single particle level is singly occupied or not, where the only difference lies in the local factor, $F_l$.   
 
To obtain $C_{p,l} \overline{ F_l}$, we first take the one-arc limit of this equation where the results are known from BCS\ theory. The one arc limit is obtained by letting $\Delta_2\to0$,  upon which $\tau\to\infty$. Taking this limit in (\ref{almostFinal}) we get a finite result only for $p=0$:
\begin{align}
\overline{\<\mbox{in};l,0|c^\dagger_{m\sigma}|\mbox{in}\>^2} \to \overline{F_l}C_{0,l}\frac{1}{2}\left[1+l\frac{\varepsilon_m-\mu_1}{\sqrt{(\varepsilon_m-\mu_1)^2+\Delta_1^2}} \right].
\end{align}
The result is true only for $l=\pm1$, while for any other $l$ the coherence factors vanish. An agreement is reached with the BCS result for:
\begin{align}\label{FlExpression}
\overline{F_l}=\frac{N_{l\sigma}}{C_{0,l}}
\end{align}
for $l=\pm1$ and $\overline{F}_l=0$ for all other $l$. 
We still need to find $C_{p,l}$ . To this end we use the following identity relating the average of  $\hat{N}_m  = c^\dagger_{m+}c_{m+} +  c^\dagger_{m-}c_{m-}$:
\begin{align}
\label{AverageAsMatrixElem}
\<{\rm in}|\overline{\hat{N}_m} |{\rm in}\> = \sum_{l,p,\sigma}\overline{\<\mbox{in};l,p|c^\dagger_{m\sigma}|\mbox{in}\>^2},
\end{align}
which relies on the fact that no other states may be excited by $c^\dagger_m$ other than those describe by $\<{\rm in};l,p|$. The left hand side of (\ref{AverageAsMatrixElem}) in known from Ref.  [\onlinecite{Gorohovsky:Bettelheim:Expectation:Values}], where
it was shown that:
\begin{align} \label{PrevExpValRes}
\left\<\overline{\frac{\hat{N}_m}{2}}\right\>  =\frac{1}{2}\left(1+\frac{\wp(u_\infty-u_{\varepsilon_m})+\wp(u_\infty+u_{\varepsilon_m})+4\frac{\eta}{\pi}}{\wp(u_\infty-u_{\varepsilon_m})-\wp(u_\infty+u_{\varepsilon_m})}\sum_\sigma\frac{N_{1,\sigma}-N_{-1,\sigma}}{2}\right)
\end{align}
Combining (\ref{PrevExpValRes}), (\ref{almostFinal}) and (\ref{FlExpression}) yields:

\begin{align}\label{ConstantsEquation1}
\sum_{lp\sigma}\frac{lN_{l\sigma}\pi^2 \sin^{-2}\left[u_{\varepsilon_m}-l(u_\infty-p \tau)\right]}{4 \omega^2 R_4\left(\varepsilon\right)}\frac{C_{p,l}}{C_{0,l}}  =
1+\frac{\wp(u_\infty-u_{\varepsilon_m})+\wp(u_\infty+u_{\varepsilon_m})+4\frac{\eta}{\pi}}{\wp(u_\infty-u_{\varepsilon_m})-\wp(u_\infty+u_{\varepsilon_m})}\sum_\sigma\frac{N_{1,\sigma}-N_{-1,\sigma}}{2}
\end{align}
The left and right hand side, as functions of $u_{\varepsilon_m}$ have poles at $\pm u_\infty + i \pi+ j\tau $ for any integer $i$ and $j$. Comparing residues gives $\frac{C_{p,l}}{C_{0,l}}=1$. Substituting  $\frac{C_{p,l}}{C_{0,l}}=1,$ equation (\ref{ConstantsEquation1}) becomes an identity as a consequence of the following known formula:
\begin{align}{  \wp(u)+\frac{2\eta}{\pi}}=\sum_{p}\sin^{-2}\left[u-p\tau\right]
\end{align}
We obtain, then, our final result,  Eq. (\ref{IntroductionResults}).
\begin{align}
&\overline {\<\mbox{in};l,p|c^\dagger_{m\sigma}|\mbox{in}\>^2}=\frac{\pi^2 l N_{l,\sigma} }{4\omega^2 R_4({\varepsilon_m})}\sin^{-2}\left[u_{\varepsilon_m}-l u_\infty+l p \tau\right]
\end{align}

\section{Acknowledgements}
We are grateful for discussions with B. Spivak and P. Wiegmann. EB is grateful for the hospitality at the University of Cologne and at the Simons Center for Geometry and Physics where this work has been completed. This work has been supported by
the Israel Science Foundation (Grant No. 852/11) and by
the Binational Science Foundation (Grant No. 2010345).

\appendix

\section{Determinant decomposition into "close" and "far" contributions \label{MatrixDecompositionAppendix}}

In order to be able to proceed, we shall need to separate out those rapidities which are close to $\varepsilon_m$ and those which are farther away. We take a large number $K$ and consider any $v_\alpha$ close to $\varepsilon_m$ if $|v_\alpha - \varepsilon_m | < K \iota$. We shall see that for large $K$, which nevertheless does not scale with $N$ (namely, $\frac{K}{N} \to 0$ as $N \to \infty$), all the determinants in question decompose into two separate factors which may be associated with the contribution of close rapidities and far rapidities, respectively. Each column, $j$, of the matrix is associated with a rapidity $w_j$. Assume that the $K$ final rows and columns of the matrix are associated with rapidities which are close to $\varepsilon_m$, while the $P-K$ first rows and columns are associated with far away rapidities.  The matrices $D(V,W;\epsilon^v), D(V,W;\epsilon^w)  $  naturally take a block matrix form:
\begin{align}\label{Edecomposition}
 D(V,W;\epsilon^v)  = \left( \begin{array}{cc}
 D_+ & \overleftrightarrow{\mathcal Y}\\
       - \overleftrightarrow{\mathcal Z} & d_+
\end{array} \right);\quad
D(W,V;\epsilon^w)  = \left( \begin{array}{cc}
 D_- & 3\overleftrightarrow{\mathcal {Y}}\\
\overleftrightarrow{\mathcal Z} & d_-
\end{array} \right).
\end{align}
where, to leading order,\begin{align}
\left[\overleftrightarrow{\mathcal Z} \right]_{i,j} = \frac{1}{2(w_i-\varepsilon_m)(w_i-w_j)} , \quad \left[\overleftrightarrow{\mathcal Y} \right]_{i,j} = \frac{1}{2(w_i-\varepsilon_m)^2}\label{YZdef}.
\end{align}
Similarly the matrices $N(W)$ , $N(V)$ take the following form to leading order:
\begin{align}
N(W)=\left( \begin{array}{cc}
N_W & 0\\
        0 & n_W
\end{array} \right),\quad N(V)=\left( \begin{array}{cc}
N_V & 0\\
        0 & n_V
\end{array} \right),
\end{align}
To write down a useful expansion for $d_\pm$, we now use the definitions for $\delta H^N(w_b),\delta H^F$ (\ref{deltaHF},\ref{deltaHN}) to write  $d_\pm$ correct to order $\iota^{-1}:$
 \begin{align}\label{DefLittlEd}
&d_{\pm a,b}=
\begin{cases}
 \frac{\delta H^{N}(w_b)^2 -\delta H'^{N}(w_b)}{2} \pm  \frac{\delta H^{N}(w_b)}{2(w_b - \varepsilon_m)}+ \delta H^F\left(\delta H^{N}(w_b)\pm\frac{1}{2(w_b-\varepsilon_m)}\right)+O(1)&a=b\\
\frac{1}{(w_a - w_b)^2} \pm \frac{1}{2(w_a - w_b)(w_a -\varepsilon_m)}&a\neq b
\end{cases}
\end{align}
Note that for $j>P-k$ the order of $\delta H^N(w_j)$\ is  $\iota^{-1}$ while $\delta H^F$ is of order $1$.
Using a general formula for the determinant of block matrices, the determinant of (\ref{Edecomposition}) can be written as
\begin{align}\label{detDecomp+}
\det D(V, W;\epsilon^v)  = \det D_+ \det d_+\det\left(I+D_{+}^{-1} \overleftrightarrow{\mathcal Y}d_+^{-1} \overleftrightarrow{\mathcal Z}\right) \\ \label{detDecomp-}
\det D(V, W;\epsilon^w)  = \det D_- \det d_-\det\left(I-3D_-^{-1} \overleftrightarrow{\mathcal Y}d_-^{-1} \overleftrightarrow{\mathcal Z}\right)
\end{align}
which may in turn be substituted into the equation for the matrix elements:\begin{align}\label{GeneralExpression}
\<\mbox{in};l,p|c^\dagger_{m\sigma}|\mbox{in}\>^2=\frac{\det\left[D(V,W;\epsilon^v)\right]\det\left[D(W,V;\epsilon^w)\right]}{\det\left[N(W)\right]\det\left[N(V)\right]}\left(1-\sum_i\vec{X}\left[D(  W,V;\epsilon^w)^{-1}\right]_i\right).
\end{align}
To obtain an expression which has a good continuum limit, we must get rid of the dependence on $d_\pm$, $\overleftrightarrow{Y}$ and $\overleftrightarrow{\mathcal Z}$. We do this in the following by either solving for the expressions involving these unwanted objects explicitly to the required order, or by separating these into factors which depend only on the local configuration of rapidities around $\varepsilon_m$ and factors which have a good continuum limit. The local factors, can then be obtained by other means, to be described in the main text. In this appendix we only concern ourselves with obtaining such a factorization into local factors and factors which have a good continuum limit.

To obtain the above mentioned factorization, we treat each of the factors going into (\ref{GeneralExpression}). The determinants $\det D_\pm$ have a good continuum limit, so we leave them as is.  The objects $D_{\pm}^{-1} \overleftrightarrow{\mathcal Y}d^{-1}_\pm \overleftrightarrow{\mathcal Z}$ appearing  in (\ref{detDecomp+}) and  (\ref{detDecomp-}), respectively, have the following representation:
\begin{align}
D_\pm^{-1} \overleftrightarrow{\mathcal Y}d_\pm^{-1} \overleftrightarrow{\mathcal Z_\pm}=D_\pm^{-1}|\vec{Y}\> \<\vec{Z}|d_\pm^{-1}|1\>\<\vec{Y}|.
\end{align}
Note that this is a  dyadic matrix. We have used here the notation $\<\vec{Y}|$ and $\<\vec{Z}|$ to denote the far elements of the vectors defined in (\ref{DefYandZ}). The determinant of the identity matrix plus a dyadic matrix is one plus the trace of the dyadic matrix. Namely, the determinants in (\ref{detDecomp+}), (\ref{detDecomp-}), which may be written as  $\det\left(I-(1\mp2)D^{(\pm)-1} \overleftrightarrow{\mathcal Y}d^{(\pm)-1} \overleftrightarrow{\mathcal Z}_\pm\right) $ can be evaluated as follows:
\begin{align}\label{DyadicAddition}
\det\left(I-(1\mp2)D_\pm^{-1} \overleftrightarrow{\mathcal Y}d_\pm^{-1} \overleftrightarrow{\mathcal Z}\right) =1-(1\mp2)\<\vec{Y}|D_\pm^{-1}|\vec{Z}\> \<\vec{Z}|d_\pm^{-1}|1\>.
\end{align}
In this last expression it remains to find a representation of $\<\vec{Z}|d_{\pm}^{-1}|1\> $, which will facilitate the derivation of the factorized form we pursue. We shall obtain this factorized form by treating the two cases $l=1$ and $l=-1$ separately, to which we devote two separate subsections below.   Before delving into the two separate cases we list the different objects which must be treated for both $l=1$ and $l=-1$:
\begin{enumerate}
\item{The determinant $\det D_\pm $,}
\item{The matrix element $  \<\vec{Z}|d_\pm^{-1}|1\> $,}
\item{The  matrix element $\<\vec{X}| D(W,V;\epsilon^w)^{-1}|1\> $. }
\end{enumerate}
\subsection{$l=1$}
In case $l=1$ we assume that $d_\pm$ to leading order is a non-singular matrix, whose leading order is given by (\ref{DefLittlEd}). All the elements of the matrix are thus local, therefore the determinant is a local factor.  Turning our attention to the second item,  $ \<\vec{Z}|d_\pm^{-1}|1\>,   $ we show that this matrix element is of  order $\iota$. Indeed, $d_\pm$ lowers the order of a vector it acts on by $2$ (namely it acts as multiplication by $\iota^{-2}$ of the vector elements). $\vec{Y}$ is of order $1$ and the summation over vector $P-k$ elements lowers the order by $1.$ In summary  $ \<\vec{Z}|d_\pm^{-1}|1\>$ is of order $\iota$, namely negligible in the continuum limit.

We are left with the third item,  $\<\vec{X}| D(V, W;\epsilon^w)^{-1}|1\> $. We define  $\vec{E}=\vec{X} D(V, W;\epsilon^w)^{-1}$, and our goal is to find $\sum_i E_i = \<\vec{X}| D(V, W;\epsilon^w)^{-1}|1\>$. we first separate out in $\vec{X}$ the  far and close parts:
\begin{align}
\vec{X} = \left( \vec{\mathcal{X}} , \vec{x} \right),\quad \vec{E} = \left(
\vec{\mathcal{E}},
\vec{e} \right)
\end{align}
which allows us to write:
\begin{align}
 \label{XInverseDAppendix} \left( \vec{\mathcal{X}} , \vec{x} \right)= \left( \begin{array}{c}
\vec{\mathcal{E }} D_- + \vec{e} \overleftrightarrow{\mathcal Z} \\
\vec{e}d_- +3 \vec{\mathcal E} \overleftrightarrow{\mathcal Y}.
\end{array}\right)
\end{align}
We now expand $\vec{X}$ and $\vec{E}$ in orders of $\iota$ defining $\vec{E} = \vec{E}^{(0)} + \vec{E}^{(1)} + \dots$ and $\vec{X} = \vec{X}^{(-1)} + \vec{X}^{(0)} + \dots$, where the superscript in parenthesis denotes the order in $\iota$.  The expression for $\vec{X}$ allows us to choose $\vec{\mathcal{X}}=\vec{\mathcal{X}}^{(-1)}+\vec{\mathcal{X}}^{(0)}$  and $ \vec{x}=\vec{x}^{(-2)}+\vec{x}^{(-1)} $, while all other terms may be taken to be zero. We shall also expand $\overleftrightarrow{\mathcal{Z}}=\overleftrightarrow{\mathcal{Z}}^{(-1)}+\overleftrightarrow{\mathcal{Z}}^{(0)}+ \dots$ and note that $\overleftrightarrow{\mathcal{Y}}$ is of order $\iota^{0}$ while $d_-$ and $D_-$ are respectively of orders $\iota^{-2}$ and $\iota^{-1}$ . We obtain the following equations by taking the $\iota^{-2}$ order of the left hand side of (\ref{XInverseDAppendix}) and collecting the terms which contribute to that order on the right hand side:
\begin{align}\label{m2Orderl1}
\left( 0 , \vec{x}^{(-2)}  \right)= \left( 
0 ,
\vec{e}^{(0)}d_{-} 
\right).
\end{align}
One can use the following identity 
\begin{align}\label{SumIntIdentity}
\sum_{i=1}^{n}F(x_i)\frac{\prod_{j=1}^n (x_i-y_j)}{\prod_{j=1(\neq i)}^n (x_i-x_j)}=\oint_{\left\{x_i\right\}_{i=1}^{n}} F(x)\frac{\prod_{j=1}^n (x-y_j)}{\prod_{j=1}^n (x-x_j)}dx
\end{align}
to show by substitution into the previous equation that:
\begin{align}
e_i^{(0)}=\left[\vec{x}^{(-2)}d_-^{-1}\right]_{i}= \frac{1}{v_i-\varepsilon_m}\frac{\prod_{j=P-k}^P (v_i-w_j)}{\prod_{j=P-k(\neq i)}^P (v_i-v_j)}\prod_{j=P-k}^P \frac{v_j-\varepsilon_m}{w_j-\varepsilon_m} +O(\iota)
\end{align}  We proceed to the next  order
\begin{align}
\left(\begin{array}{c} \vec{\mathcal{X}}^{(-1)} \\ \vec{x}^{(-1)} \end{array} \right)= \left( \begin{array}{c}
  \vec{\mathcal{E}}^{(0)} D_{-}+\vec{e}^{(0)} \overleftrightarrow{\mathcal Z}^{(-1)}     \\
\vec{e}^{(1)}d_{-} +3\vec{\mathcal{E}}^{(0)} \overleftrightarrow{\mathcal Y}   
\end{array}\right).\label{m1leq1}
\end{align}   
We find $\vec{e}^{(0)} \overleftrightarrow{\mathcal Z}^{(-1)} =  \vec{\mathcal{X}}^{(-1)}$ such that we have $ \vec{\mathcal{E}}^{(0)} = 0$ and thus, using (\ref{SumIntIdentity}):
\begin{align}
&e_i^{(1)} =\left[\vec{x}^{(-1)}d_-^{-1}\right]_i=\frac{\prod_{j=P-k}^P (v_i-w_j)}{\prod_{j=P-k(\neq i)}^P (v_i-v_j)}\prod_{j=P-k}^P \frac{v_j-\varepsilon_m}{w_j-\varepsilon_m}\delta H^F.
\end{align}
We shall also require $\vec{\mathcal{E}}^{(1)}$ which is determined by the following $\iota^0$ order equation:
\begin{align}
\vec{\mathcal{X}}^{(0)} =\vec{\mathcal{E}}^{(1)}D_{-}+ \vec{e}^{(1)}\overleftrightarrow{\mathcal Z}^{(-1)}   + \vec{e}^{(0)}\overleftrightarrow{\mathcal Z}^{(0)},   
\end{align} 
where we use the subleading order of  $\overleftrightarrow{\mathcal Z}$, given by $\overleftrightarrow{\mathcal Z}^{(0)}=\frac{1}{(w_i-w_j)^2}$.
Upon substitution it can be shown that the solution of the previous equation is:
\begin{align}
\vec{\mathcal{E}}^{(1)}=\vec{\mathcal{X}}^{(0)} \left[D_{-}\right]^{-1} \prod_{j=P-k}^P \frac{v_j-\varepsilon_m}{w_j-\varepsilon_m}
\end{align}
Finally using the expressions for $\vec{\mathcal{E}}^{(1)}, \vec{e}^{(0)}$ together with (\ref{SumIntIdentity}) one finds:
\begin{align}
\sum_i E_i=1-\left(1-\left\<\vec{\mathcal{X}}^{(0)}\left| \left[D_{-}\right]^{-1}\right|1\right\>\right) \prod_{j=P-k}^P \frac{v_j-\varepsilon_m}{w_j-\varepsilon_m}
\end{align}
The final expression for the overlap in case $l=1$ is then:
\begin{align}\label{DecompositionCaseI}
&\<\mbox{in};1,p|c^\dagger_{m\sigma}|\mbox{in}\>^2=\frac{ \det\left[ D_{+}D_{-}\right] }{ \det \left[N_W N_V\right]}\left(1-\left\<\vec{\mathcal{X}}^{(0)}\left|[D^{-}]^{-1}\right|1\right\>\right)F_{1}
\end{align}
where
\begin{align}
\mathcal{X}_i^{(0)}=\frac{1}{2(w_i-\varepsilon_m)^{2}}-\frac{\delta H^{F}}{2(w_i-\varepsilon_m)};\quad F_{1}=\frac{ \det \left[d_{+} d_{-} \right]}{\det \left[n_W n_V\right]}\prod_{j=P-k}^P \frac{v_j-\varepsilon_m}{w_j-\varepsilon_m}
\end{align}

\subsection{$l=-1$}
It turns out that, for $l=-1$, the leading order to the  matrix $d_+$ has a null vector, namely a vector with $O(1)$ elements which satisfies:
\begin{align}
\<\vec{p}|d_+  =0 + O(\iota^{-1}).
\end{align}
Note that  $d_+$ is of order $\iota^{-2}$ such that if $\vec{p}$ were not a null vector then $\vec{p}d_+ = O(\iota^{-2})$. We write an explicit expression for $\vec{p}$:
\begin{align}
p_i=\frac{\prod_{j>P-K}^P (w_i-v_j)}{\prod_{j>P-K} (w_i-w_j)}
\end{align}
where the fact that $\vec{p}$ is indeed a null vector can be verified by direct substitution. This means that $d_+$ is singular in leading order, the effect of this singularity is to make the magnitude of  $\<\vec{Z}|d_+^{-1}|1\>$  to be $O(1)$ as  opposed to the  $l=1$  case, where it was negligible. In addition to this, one can use (\ref{DefLittlEd}) to show that in leading order:
\begin{align}\label{PropertiesDIIB+}
&\<\vec{Z}|d_+^{-1}=\frac{1}{\delta H^F}\left\<\vec p \right|+O(1)
\end{align}
We can now use (\ref{PropertiesDIIB+}) and $\<\vec{p}|\vec {1}\>=1$  to perform the decomposition of $\det D(V, W;\epsilon^v) $, in leading order, we have:
\begin{align}
\det D(V, W;\epsilon^v) & \label{DpDecomleqm1}= \det D_{+} \det d_{+}\left(1+\left\<\vec{Z}\left|d_{+}^{-1}\right|\vec 1\right\>
\left\<\vec{Z}\left|D_{+}^{-1}\right|\vec{Y}\right\>\right)\nonumber\\
&=\det D_{+} \frac{\det d_{+}}{\delta H^F}\left(\delta H^F+\left\<\vec{Z}\left|D_{+}^{-1}\right|\vec{Y}\right\>\right)
\end{align}
The dependence of $\det d_{+}$ on far rapidities is manifested only in sub-leading order which is proportional to  $\delta H^F$  (\ref{DefLittlEd}).  Since $d_{+}$ is singular to leading order, the determinant a product of $\delta H^{F}$ and some function dependent on the near rapidities. This means that although, $\frac{\det d^{+}}{\delta H^F}$ is a local factor. Since $\<\vec{Z}|d_-^{-1}|1\>$, it negligible,  the decomposition of $\det D(V, W;\epsilon^w) $ has the same for as  $l=1$, i.e
\begin{align}
\det D(V, W;\epsilon^v) \label{D-Decompleqm1}= \det D_{-} \det d_{-}
\end{align}
To complete the calculation we must find $\<\vec{X}| D(V, W,\epsilon^w)^{-1}|1\> $.  We solve find $\<\vec{X}| D(V, W,\epsilon^w)^{-1} $ by expanding the equation     $\vec{E}=\vec{X} D(V, W;\epsilon^w)^{-1}$ order by order just as the case $l=1$. The first equation we encounter is (\ref{m2Orderl1}). Which is solved by $ \vec {e}^{(0)}=\vec{x}^{(-2)}d_-^{-1}$. Since both $d_-$ and $\vec{x}^{(-2)}$ are local to leading order, we conclude that the elements of $\vec{e}^{(0)}$ are local. 

Next we find (\ref{m1leq1}). Which suggests:
\begin{align}\label{WSolutionCaseII1}
\mathcal{\vec X}^{(-1)}=\vec{\mathcal{E}}^{(0)}D_{-}+e^{(0)}\mathcal{Z}
\end{align}
or
\begin{align}\label{WSolutionCaseII}
\vec{\mathcal{E}}^{(0)}=\left(\vec{\mathcal{X}}^{(-1)}-\vec{e}^{(0)}\mathcal{Z}^{(-1)}\right)D_{-}^{-1}
\end{align}
Using the explicit form of $\vec{X}, $ Eq. (\ref{DefBigX}), and $\mathcal{Z}$, Eq. (\ref{YZdef}), one can write $\vec{\mathcal{X}}^{(-1)} = -\delta H^N(\varepsilon_m)\<\vec{Z}|$,   and $\vec{\mathcal{Z}}^{(-1)} = |\vec{z}\>\<\vec{Z}|$, where here $|\vec{Z}\>$ denotes the far elements of $\vec{Z}$ as defined in (\ref{DefYandZ}) and $|\vec{z}\>$ the close elements.  This allows us to write:
\begin{align}
\vec{\mathcal{E}}^{(0)} =-\left(\delta H^N  (\varepsilon_m)+ \<\vec{e}^{(0)}|\vec{z}\> \right)\<\vec{Z}|D_-^{-1}
\end{align}
and thus to leading order:
\begin{align}
\label{Cramersleqm1}\<\vec{X}| D(V, W,\epsilon^w)^{-1}|1\>=-\left(\delta H^N(\varepsilon_m)+\<\vec{e}^{(0)}|\vec{z}\> \right)\left\<\vec{Z}\left|D_{-}^{-1}\right|1\right\>
\end{align}
Combining (\ref{DpDecomleqm1})  (\ref{D-Decompleqm1})  and (\ref{Cramersleqm1}), we obtain: 
\begin{align}\label{DecompositionCaseII}
&\<\mbox{in};-1,p|c^\dagger_{m\sigma}|\mbox{in}\>^2=\frac{ \det \left[D_{+}D_{-}\right] }{ \det \left[N_W N_V\right]}\left(\delta H^F+\left\<\vec{Z}\left|D_{+}^{-1}\right|\vec{Y}\right\>\right)\left\<\vec{Z}\left|D_{-}^{-1}\right|1\right\>F_{-1}
\end{align}
where
\begin{align}
&F_{-1}=\left(\delta H^N(\varepsilon_m)+\sum_{i=P-k+1}^{P}Z_i e^{(0)}_i\right)\frac{ \det d_{+}}{\delta H^F}\frac{ \det d_{-}}{\det \left[n_Wn_V\right]},
\end{align}
is a local factor.

\section{Derivation of Eq. (\ref{DactionLineIntegral}) \label{KSumsAppendix}}
We are interested in proving (\ref{DactionLineIntegral}), for that purpose we must make and explicit calculation of $ \sum_{|s(a) - s|<r} D_{+a , i(s)}$ in (\ref{xOFsSumIntegral}) up till the sub leading order $\iota^{-1}$. The diagonal element of $D_+$   is features $\delta H^{}$ which is given by
\begin{align}
\delta H^{}(w_{a(s)})=\sum_{i\neq a(s)}\frac{1}{w_{a(s)}-w_i}-\sum_{i}\frac{1}{w_{a(s)}-v_i}
\end{align}
We now separate each sum in to two sums that run over rapidities close and far from $w_a$ . The far part is not affected by the exact position of $w_b$ and we can evaluate it from either sides of the arc, this allows us to exploit a continuous property of Richardsons solution to show that
\begin{align}\label{ContinuousProp}
\delta H^F(w_{a(s)})=\left[\delta h^{F}(s(a))-\frac{1}{w_{a(s)}-\varepsilon_m}\right]^A=-\frac{1}{w_{a(s)}-\varepsilon_m}
\end{align}
We compute now the contribution from the close elements.  To evaluate the off diagonal sum, we first write an expression for the distance of $w_a(s)$ to its close neighbor $w_{a(s)+k}$ as follows
\begin{align}
w_{a(s)}-w_{a(s)+k}&=(w_{a(s)}-w_{a(s)+1})+(w_{a(s)+1}-w_{a(s)+2})...+(w_{a(s)+k-1}-w_{a(s)+k})\nonumber\\
&\approx\frac{k\iota  }{\sigma(s)}+\sum_{i=1}^{k-1}\frac{\sigma'(s)\iota^2 }{\sigma(s)^3}i=\frac{k\iota  }{\sigma(s)}+\frac{\sigma'(s)k\iota^2 }{\sigma(s)^3}\frac{(k-1)k}{2}
\end{align}
where $\frac{\sigma(s)}{\iota}$ is the density of the rapidities on the arc. Similarly
\begin{align}
w_{a(s)}-v_{a(s)+k}=\frac{k\iota  }{\sigma(s)}+\frac{\sigma'(s)k\iota^2 }{\sigma(s)^3}\frac{(k-1)k}{2}-\left(\iota\delta w(s)+\frac{\delta w(s)'\iota^2 k}{\sigma(s)}\right)
\end{align}
where $\iota\delta w(s)$ is the relative shift between $W$ and $V$ and has an order of $\iota$.
We can now use these equations to perform the summation over the close rapidities correct to leading order:
\begin{align}\label{DeltaHCloseExp}
\delta H^{N}(s)&=\sum_{k(\neq 0)=-\infty}^\infty\left(\frac{\sigma(s)}{k\iota}+\frac{\sigma(s)'(k-1)}{ \sigma(s)2k}\right)-\sum_{k=-\infty}^\infty\left(\frac{\sigma(s)}{k\iota-\iota\delta w(s)\sigma(s)}+\frac{\sigma(s)'(k-1)+2\sigma(s)^2\delta w(s)'}{ 2\sigma(s)\left(k-\delta w(s)\sigma(s)\right)^2}k\right)\nonumber\\
&=-\frac{\pi \sigma(s)}{\iota}\cot\left(\pi \sigma(s)\delta w(s)\right)
\end{align}
Similarly, for the derivative , $\delta H'^{N}(s),$ we find
\begin{align}\label{DeltaH'CloseExp}
&\delta H^{'N}(s)\approx\sum_{k(\neq 0)=-\infty}^\infty\left(\frac{\sigma(s)^2}{k^2\iota^2}+\frac{\sigma(s)'(k-1)}{ k^2\iota}\right)-\sum_{k=-\infty}^\infty\left(\frac{1}{\iota^2\left(\delta w(s)-\frac{k}{\sigma(s)}\right)^2}+\frac{\sigma(s)'(k-1)+2\sigma(s)^2\delta w(s)'}{\iota\left(k-\sigma(s)\delta w(s)\right)^3}k\right)\nonumber\\
&=\frac{\pi^2 \sigma(s)^2}{\iota^2}\left(\cot\left(\pi \sigma(s)\delta w(s)\right)^2+\frac{2}{3}\right)+\frac{2\pi^2\sigma(s)}{\iota}\left(\sigma(s)\delta w(s)\right)'-\frac{2\pi ^2\sigma(s)' }{3\iota}
\end{align}
We now substitute (\ref{ContinuousProp},\ref{DeltaHCloseExp},\ref{DeltaH'CloseExp}) into the definition of the diagonal term of $D_+$ (\ref{OverlapsAsDdeterminant}) to obtain
\begin{align}\label{TheDiagonalElementOfD1}
D_{+\,\,b(s),b(s)}=-\frac{\pi^2 \sigma(s)^2}{3\iota^2}+\frac{\pi ^2\sigma(s)' }{3\iota}+\frac{\pi^2\sigma(s)}{\iota}\left(\sigma(s)\delta w(s)\right)'
\end{align}
For the sum of the off diagonal elements we have
\begin{align}
 \sum_{|s(a) - s|<r\neq 0,s} D_{+\,\,a , i(s)}=\sum_{k(\neq 0)=-\infty}^\infty\left(\frac{\sigma(s)^2}{k^2\iota^2}+\frac{\sigma(s)'(k-1)}{ k^2\iota}\right)=\frac{\pi^2 \sigma(s)^2}{3\iota^2}-\frac{\pi ^2\sigma(s)' }{3\iota}
\end{align}
such that the overall sum is
\begin{align}
 \sum_{|s(a) - s|<r} D_{+\,\,a , i(s)}=\frac{\pi^2\sigma(s)}{\iota}\left(\sigma(s)\delta w(s)\right)'
\end{align}
Similarly, one can find the diagonal element and the overall sum of $D_-$ by taking $\delta w(s)\rightarrow-\delta w(s)$ and of $N(W)$ by taking $\delta w(s)\rightarrow 0$.

\section{Determinants of Matrices as Determinants of Operators  \label{CanTakeDetOfOperatorsAppendix}}
In this Appendix, our goal is to show how to find a continuum limit representation of :
\begin{align}
d_{\varepsilon} \log \frac{\det \left[D_+ D_- \right]}{\det \left[ N_WN_V \right]}= \tr \left[ D_{+}^{-1} d_{\varepsilon}\delta D_++D_-^{-1} d_{\varepsilon}\delta D_- -\left(N_V^{-1} -D_-^{-1}\right) d_{\varepsilon}N_V-\left(N_W^{-1} -D_+^{-1}\right) d_{\varepsilon}N_W\right],
\end{align}
 where $\delta D_+=D_+-N_V$ and $\delta D_- = D_--N_W$. We note that to leading order $\delta D_+ = -\delta D_-$ such that we may write:
\begin{align}
 \label{ToolongTrace}d_{\varepsilon} \log \frac{\det \left[D_+ D_- \right]}{\det \left[ N_WN_V \right]}= \tr \left[ \left(D_{+}^{-1} -D_-^{-1} \right)d_{\varepsilon}\delta D_+ -\left(N_V^{-1} -D_-^{-1}\right) d_{\varepsilon}N_V-\left(N_W^{-1} -D_+^{-1}\right) d_{\varepsilon}N_W \right],
\end{align}

 We shall show that the rows  of each of the matrices $\left(D_{+}^{-1} -D_-^{-1} \right)$, $\left(N_V^{-1} -D_-^{-1}\right)$ and $\left(N_W^{-1} -D_+^{-1}\right)$ are smoothly varying vectors, a fact which allows us to replace all matrices by their continuum limit.  Noting that in the continuum limit $N_V$ and $N_W$ are described by the operator $\partial_\xi $, we see that the derivative of these matrices may be dropped and we obtain to leading order :

\begin{align}\label{TraceExpression}
 d_{\varepsilon} \log \frac{\det \left[D_+ D_- \right]}{\det \left[ N_WN_V \right]}= \tr \left[ \left( \mathcal{D}_{+}^{-1} -\mathcal{D}_-^{-1} \right)d_{\varepsilon} \mathcal{D}_+   \right].
\end{align}

We now turn to show that  rows of the appropriate matrices are indeed smoothly varying. Let  $S$
  stand for one of the matrices $D_+$, $D_-$, $N_V$ or $N_W$. We choose $1\ll K\ll P, $ fix some $j$ and look at the $j$'s row of $S^{-1}$. Within this row we separate the elements close to the diagonal, $\left\{S_{ ij}^{-1};|i-j|<K\right\}$, and  those far from it, $\left\{S_{ ij}^{-1};|i-j|\geq K\right\}$. The far part is well approximated by solving the equations:
\begin{align}\label{gContDEF}
\mathcal{S} g^S_{v_j}(\xi) = \frac{1}{\xi -v_j}
\end{align}
and taking the jump discontinuity at $v_i$:
\begin{align}
\label{FarDMinus1Approx}S_{ ij}^{-1} \simeq \frac{\iota^2\left( g_{v_j}^{S}\right)^{ J}(v_i)}{4\pi^2\sigma(i) \sigma(j)}.
\end{align}
where $\mathcal{S}$ is the continuum limit of $S$, i.e. $\mathcal{D}_\pm$ or $\partial_\xi$. Indeed, the right hand side of (\ref{gContDEF}) is the representation of the vector $v_i = \delta_{i,j}$ and thus (\ref{gContDEF}) is the operator representation of the matrix equation $\sum_k S_{ i,k} S^{-1}_{ k,j} = \delta_{i,j}$ where $g^S_{v_j}(\xi)$ is the continuum limit representation of the $j$'s column of $S^{-1}$.

One may write down an approximation for the matrix $S_{ ab}$ valid for $|a-j|<K$ and $|b-j|<K$. We denote this approximation by $S^N$:
\begin{align}
S_{ a+j,b+j} \simeq S^N_{a,b}.
\end{align}
The crucial point is that all the matrices for which $S$ stand have the same approximation:
\begin{align}\label{CloseSolution} S_{ab}^N =
-\frac{2\pi^2\sigma(j)^2}{\iota^{2}}\left\{ \begin{array}{lr}
-\frac{\pi^2}{3}&a=b \\
\frac{1}{i^2}&a\neq b
\end{array}\right. .
\end{align}
Note that the matrix indices in $S^N$ run from $-K$ to $K$. We now take $K$\ to infinity. The infinite matrix $S^N$ does not have an inverse since the constant vector $|1\>$ is a zero mode $S^N|1\>=0$. The equation
\begin{align}
S^N_{ a,b}g_b =\delta_{a,0}
\end{align}
has a solution which tends to at infinity. The solution may be written explicitly, since the matrix depends only on the difference of the indices $S^N_{ij} = S^N_{i-j}$ and as such equations involving it may be solved by Fourier transform. We omit this here. We approximate $S^{-1}_{ij}$ by:
\begin{align}\label{CloseDmis1Approx}
S^{-1}_{ij} \simeq g_{i-j} + c_S ,
\end{align}
where  only the constant, $c_S$, depends on the identity of the matrix $S$, and   must be determined by matching the approximation (\ref{CloseDmis1Approx}) to the far approximation (\ref{FarDMinus1Approx}) in the region where both are applicable, namely, $1\ll |i-j| \ll P$.  In such a region $g_{i-j}$ tends to zero such that $c_S$ is given by
\begin{align}
\label{LastEq}c_S =\frac{\iota}{2\pi \sigma(j) } \lim_{\xi \to v_j} \left(g^{S}\right)^{ J} _{v_j}(\xi)
\end{align}

Returning to the rows of the matrices   $\left(D_{+}^{-1} -D_-^{-1} \right)$, $\left(N_V^{-1} -D_-^{-1}\right)$ and $\left(N_W^{-1} -D_+^{-1}\right)$ , we note that we may use (\ref{CloseDmis1Approx}) to obtain an approximation for  elements near the diagonal. The the rapidly changing part, $g,$ cancels out, and only the slowly varying part, which around the diagonal is described by the constant $c_S$ remains. Away from the diagonal and approximation for the elements is given by solving (\ref{gContDEF}), which smoothly meshes with the constant $c_S$ near the diagonal due to (\ref{LastEq}).   
\bibliographystyle{apsrev4-1}
\bibliography{mybib}
\end{document}